\documentclass[useAMS,usenatbib]{mnras}

\usepackage{graphicx}
\usepackage{graphics}
\usepackage{color}
\usepackage{url}
\usepackage[caption=false]{subfig}
\usepackage{verbatim}
\usepackage{array}
\usepackage{amssymb}
\usepackage{amsmath}
\usepackage{longtable}
\usepackage{pbox}

\newcommand{\name}{ASASSN-15oi}
\newcommand{\galname}{2MASX J20390918-3045201}
\newcommand{\swift}{{\it Swift}}
\newcommand{\msun}{\ensuremath{\rm{M}_\odot}}
\newcommand{\lsun}{\ensuremath{\rm{L}_\odot}}

\newcolumntype{C}[1]{>{\centering\arraybackslash}m{#1}}
\newcolumntype{L}[1]{>{\arraybackslash}m{#1}}

\def\apjl{\rm ApJL}
\def\apjs{\rm ApJS}

\def\aap{\rm AAP}

\begin{document}

\title[The Luminous TDE ASASSN-15oi]{ASASSN-15oi: A Rapidly Evolving, Luminous Tidal Disruption Event at 216 Mpc}

\author[T.~W.-S.~Holoien et al.]{T.~W.-S.~Holoien$^{1,2,3}$, C.~S.~Kochanek$^{1,2}$,  J.~L.~Prieto$^{4,5}$, D.~Grupe$^{6}$, Ping Chen$^{7}$, 
\newauthor
D.~Godoy-Rivera$^{1}$, K.~Z.~Stanek$^{1,2}$, B.~J.~Shappee$^{8,9}$, Subo~Dong$^{10}$, J.~S.~Brown$^{1}$,
\newauthor
U.~Basu$^{1,11}$, J.~F.~Beacom$^{1,2,12}$, D.~Bersier$^{13}$, J.~Brimacombe$^{14}$, E.~K.~Carlson$^{8}$, 
\newauthor
E.~Falco$^{15}$, E.~Johnston$^{15}$, B.~F.~Madore$^{8}$, G.~Pojmanski$^{17}$, and M.~Seibert$^{8}$ \\ \\
  $^{1}$ Department of Astronomy, The Ohio State University, 140 West 18th Avenue, Columbus, OH 43210, USA \\
  $^{2}$ Center for Cosmology and AstroParticle Physics (CCAPP), The Ohio State University, 191 W. Woodruff Ave., Columbus, OH 43210, USA \\
  $^{3}$ US Department of Energy Computational Science Graduate Fellow \\
  $^{4}$ N\'ucleo de Astronom\'ia de la Facultad de Ingenier\'ia, Universidad Diego Portales, Av. Ej\'ercito 441, Santiago, Chile \\
  $^{5}$ Millennium Institute of Astrophysics, Santiago, Chile \\
  $^{6}$ Department of Earth and Space Science, Morehead State University, 235 Martindale Dr., Morehead, KY 40351, USA \\
  $^{7}$ Department of Astronomy, Peking University, Yi He Yuan Road 5, Hai Dian District, Beijing 100871, China \\
  $^{8}$ Carnegie Observatories, 813 Santa Barbara Street, Pasadena, CA 91101, USA \\
  $^{9}$ Hubble and Carnegie-Princeton Fellow\\
  $^{10}$ Kavli Institute for Astronomy and Astrophysics, Peking University, Yi He Yuan Road 5, Hai Dan District, Beijing 100871, China \\
  $^{11}$ Grove City High School, 4665 Hoover Road, Grove City, OH 43123, USA \\
  $^{12}$ Department of Physics, The Ohio State University, 191 W. Woodruff Ave., Columbus, OH 43210, USA \\
  $^{13}$ Astrophysics Research Institute, Liverpool John Moores University, 146 Brownlow Hill, Liverpool L3 5RF, UK \\
  $^{14}$ Coral Towers Observatory, Cairns, Queensland 4870, Australia \\
  $^{15}$ Harvard-Smithsonian Center for Astrophysics, 60 Garden St., Cambridge, MA 02138, USA \\
  $^{16}$ European Southern Observatory, Alonso de Cord\'ova 3107, Casilla 19001, Santiago, Chile \\
  $^{17}$ Warsaw University Astronomical Observatory, Al. Ujazdowskie 4, 00-478 Warsaw, Poland
  }
  
\maketitle

\begin{abstract}
We present ground-based and {\swift} photometric and spectroscopic observations of the tidal disruption event (TDE) {\name}, discovered at the center of {\galname} ($d\simeq216$~Mpc) by the All-Sky Automated Survey for SuperNovae (ASAS-SN). The source peaked at a bolometric luminosity of $L\simeq1.3\times10^{44}$~ergs~s$^{-1}$ and radiated a total energy of $E\simeq6.6\times10^{50}$~ergs over the first $\sim3.5$ months of observations. The early optical/UV emission of the source can be fit by a blackbody with temperature increasing from $T\sim2\times10^4$~K to $T\sim4\times10^4$~K while the luminosity declines from $L\simeq1.3\times10^{44}$~ergs~s$^{-1}$ to $L\simeq2.3\times10^{43}$~ergs~s$^{-1}$, requiring the photosphere to be shrinking rapidly. The optical/UV luminosity decline during this period is most consistent with an exponential decline, $L\propto e^{-(t-t_0)/\tau}$, with $\tau \simeq46.5$~days for $t_0\simeq57241.6$ (MJD), while a power-law decline of $L\propto (t-t_0)^{-\alpha}$ with $t_0\simeq57212.3$ and $\alpha=1.62$ provides a moderately worse fit. {\name} also exhibits roughly constant soft X-ray emission that is significantly weaker than the optical/UV emission. Spectra of the source show broad helium emission lines and strong blue continuum emission in early epochs, although these features fade rapidly and are not present $\sim3$ months after discovery. The early spectroscopic features and color evolution of {\name} are consistent with a TDE, but the rapid spectral evolution is unique among optically-selected TDEs. 
\end{abstract}
\begin{keywords}
accretion, accretion disks --- black hole physics --- galaxies: nuclei
\end{keywords}


\section{Introduction}
\label{sec:intro}

When a star falls deep enough into the gravitational potential of a supermassive black hole (SMBH), the tidal shear forces overpower the self-gravity of the star, resulting in a tidal disruption event (TDE). A portion of the disrupted stellar material is ejected while a fraction of the remaining material is accreted onto the black hole, creating a luminous, short-lived ($t\la1$~yr) accretion flare \citep[e.g.,][]{lacy82,rees88,evans89,phinney89}. When the central black hole has a mass $M_{BH}\la10^7 {\msun}$, the debris initially falls back onto the black hole at a super-Eddington rate, and the debris eventually returns to pericenter on a timescale that roughly follows a $t^{-5/3}$ power law for a main sequence star \citep{evans89,phinney89}. It was commonly assumed that the luminosity of the TDE flare would be proportional to this rate of return of the stellar material to pericenter and that it would peak at soft X-ray energies.

Observed TDE candidates have exhibited a large diversity in properties. TDE flares have now been discovered at hard X-ray \citep[e.g.,][]{bloom11,burrows11,cenko12b,pasham15}, soft X-ray \citep[e.g.,][]{bade96,grupe99,komossa99,donley02,maksym10,saxton12}, ultraviolet \citep[e.g.,][]{stern04,gezari06,gezari08,gezari09}, and optical \citep[e.g.,][]{velzen11,gezari12b,cenko12a,arcavi14,chornock14,holoien14b,vinko15,holoien16a} wavelengths---see \citet{komossa15} for a review. While some TDEs appear to follow the predicted $t^{-5/3}$ power law decline, others exhibit different decline rates \citep[e.g.,][]{vinko15,holoien16a,brown16}. Many of the X-ray-bright sources do not show associated strong optical emission, and many of the recent candidates discovered by optical surveys show strong optical and UV emission without associated X-ray emission. The notable exception to this is the exceptionally well-studied TDE ASASSN-14li, which was discovered in the optical \citep{holoien16a} but also detected at both X-ray \citep{holoien16a,miller15} and radio wavelengths \citep{alexander15,velzen15}. 

In practice, the properties of TDE flares depend on numerous physical parameters, including the depth of the encounter, the composition of the star, the fraction of the star that is accreted, and the geometry of the accretion stream \citep[e.g.,][]{kochanek94,lodato11,guillochon15,metzger15,shiokawa15}. If the TDE flare is powered by the accretion of the stellar material onto the black hole, the observed optical/UV emission is likely dominated by a photosphere formed within the stellar debris \citep{evans89,loeb97,ulmer99,strubbe09,roth15}, and is likely dependent on the viewing angle \citep{guillochon14,metzger15}. If the flare is powered by interactions within the debris stream during disk formation, the observed emission depends strongly on the dynamics of the system, the outflow of unbound material, and on the orientation of the debris with respect to the observer \citep{piran15,shiokawa15,svirski15}.

The optically discovered TDEs also exhibit a continuum of spectroscopic properties ranging from He-dominated to H-dominated \citep{arcavi14}. This diversity could be due to photoionization physics \citep[e.g.,][]{guillochon14,gaskell14,roth15}, composition variations created by stellar initial composition and evolution \citep{kochanek15} and/or require the disruption of some helium stars \citep{gezari12b,strubbe15}. The discovery and observation of additional nearby TDEs are needed to fully understand the processes governing the observed emission of these optically bright TDE flares.

Here we describe the discovery and follow-up observations of {\name}, the third TDE discovered by the All-Sky Automated Survey for SuperNovae (ASAS-SN\footnote{\url{http://www.astronomy.ohio-state.edu/~assassin/}}; \citealt{shappee14}). ASAS-SN is a long-term project designed to monitor the entire sky on a rapid cadence to find nearby supernovae \citep[e.g.,][]{dong16,holoien15a,shappee15} and other bright transients. In particular, ASAS-SN has already proven to be a powerful tool for discovering new TDEs, and has produced the two closest TDEs ever discovered at optical wavelengths \citep{holoien14b,holoien16a}. Our transient detection pipeline was triggered on 2015 August 14 by a new source at RA/Dec $=$ 20:39:09.12/$-$30:45:20.84 (J2000) with $V=16.2\pm0.1$~mag \citep{asassn15oi_atel}. The source was not detected ($V\ga17.2$~mag) on 2015 July 26 or earlier. The object's position corresponds to {\galname}, a galaxy with no previous redshift measurement. Follow-up images obtained on 2015 August 25 with the {\swift} UltraViolet and Optical Telescope \citep[UVOT;][]{roming05} and X-ray Telescope \citep[XRT;][]{burrows05} confirmed the detection of the transient in both the UV and X-rays. 


\begin{figure*}
\begin{minipage}{\textwidth}
\centering
\subfloat{{\includegraphics[width=0.98\textwidth]{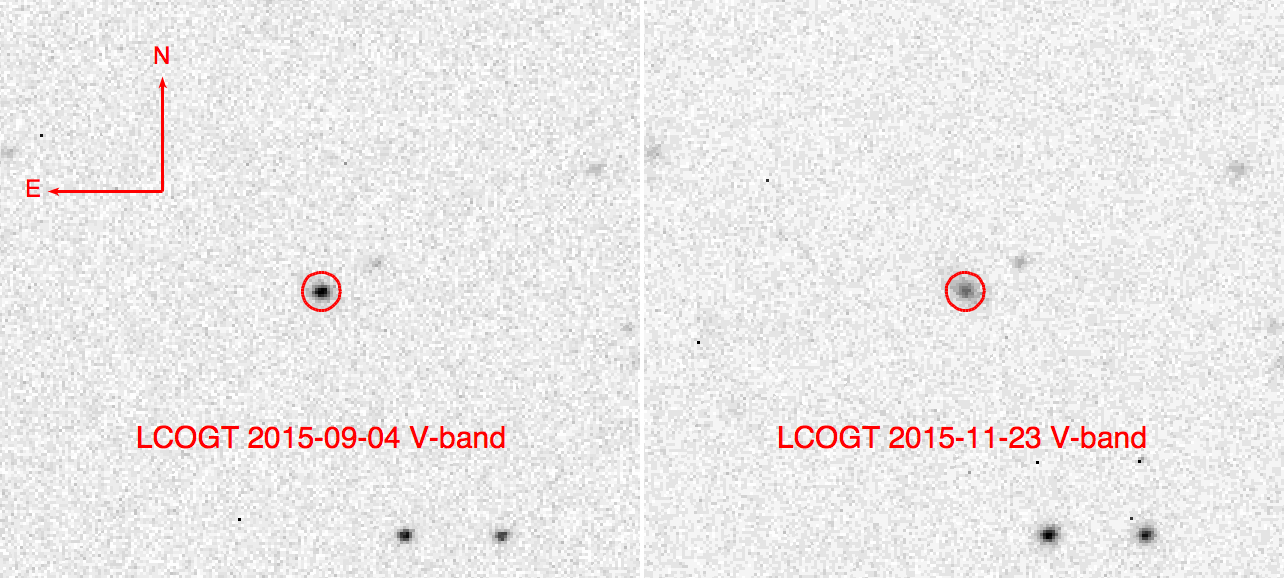}}}
\caption{$V$-band images of {\name} shortly after discovery and after the transient had faded. The left panel shows the LCOGT 1-m image from 2015 September 4 and the right panel shows the LCOGT 1-m image from 2015 November 23. The red circles have radii of 3\farcs{0} and are centered on the position of the transient measured in the September 4 image.}
\label{fig:finding_chart}
\end{minipage}
\end{figure*}

A follow-up spectrum obtained by the Public ESO Spectroscopic Survey for Transient Objects \citep[PESSTO;][]{smartt15} using the ESO New Technology Telescope at La Silla showed that the transient exhibited a strong blue continuum and a broad ($\sim10000$~km~s$^{-1}$) \ion{He}{ii} 4686\AA~emission feature \citep{asassn15oi_spec_atel}, consistent with helium-rich optical TDEs \citep[e.g.,][]{gezari12b,arcavi14}. Although the redshift was originally reported as $z=0.02$, our later spectra showed that the redshift of the host galaxy is $z=0.0484$, corresponding to a luminosity distance of 216 Mpc ($H_0=69.6$~km~s$^{-1}$~Mpc$^{-1}$, $\Omega_M=0.29$, $\Omega_{\Lambda}=0.71$). Because of the similarities between {\name} and previously discovered TDEs, we began a long-term follow-up campaign to monitor and characterize the object.

In \S\ref{sec:obs} we describe pre-outburst data of the host galaxy as well as new observations obtained of the transient during our follow-up campaign. In \S\ref{sec:analysis} we analyze these data to model the transient's luminosity and temperature evolution and compare the properties of {\name} to those of supernovae and other TDEs to determine the nature of the transient. Finally, in \S\ref{sec:disc} we discuss our findings and how they relate to the active field of TDE research.


\section{Observations and Survey Data}
\label{sec:obs}

Here we summarize the available archival survey data of the transient host galaxy {\galname} as well as our new photometric and spectroscopic observations of {\name}.


\subsection{Archival Data}
\label{sec:arch_dat}

Because the source is located in the Southern hemisphere, there are no available archival imaging or spectroscopic data in the Sloan Digital Sky Survey (SDSS). We obtained archival near-IR $JHK_s$ images from the Two-Micron All Sky Survey \citep[2MASS;][]{skrutskie06} and measured 5\farcs{0} aperture magnitudes for the galaxy in these images. This aperture radius, chosen to match the PSF of the {\swift} data, was also used to measure the source flux in follow-up data. We also obtained a $V$-band aperture magnitude for the host galaxy by stacking several epochs of ASAS-SN data and measuring the magnitude in the same way using a 2-pixel aperture, roughly corresponding to a 14\farcs{0} aperture due to the large ASAS-SN pixel scale. These measured magnitudes were later used to model the host galaxy SED and subtract the host galaxy flux from follow-up data of the transient. We present the measured 5\farcs{0} aperture magnitudes from the 2MASS and ASAS-SN images in Table~\ref{table:host_mags}.


\begin{table}
\centering
\caption{Archival Photometry of the Host Galaxy}
\label{table:host_mags}
\begin{tabular}{@{}ccc}
\hline
Filter & Magnitude & Magnitude Uncertainty \\
\hline
$NUV$ & $>20.57$ & --- \\
$V$ & 17.13 & 0.05 \\
$J$ & 15.17 & 0.07 \\
$H$ & 14.52 & 0.07 \\
$K_s$ & 14.05 & 0.09 \\
$W1$ & 14.18 & 0.03 \\
$W2$ & 14.12 & 0.05 \\
\hline
\end{tabular}

\medskip
\raggedright
\noindent 
Measured 5\farcs{0}-radius aperture $JHK_s$ magnitudes from 2MASS and a 14\farcs{0}-radius aperture $V$-band magnitude from ASAS-SN. The WISE $W1$ and $W2$ are PSF photometry magnitudes from the AllWISE source catalog. The GALEX $NUV$ limit is a 3-sigma upper limit measured from co-added exposures totaling $\sim200$s of exposure time. All magnitudes are in the Vega system.
\end{table}

{\galname} is not detected in archival Spitzer, Herschel, Hubble Space Telescope (HST), Chandra, X-ray Multi-Mirror Mission (XMM-Newton), NRAO VLA Sky Survey (NVSS), Sydney University Molonglo Sky Survey (SUMSS), or Very Large Array Faint Images of the Radio Sky at Twenty-cm (VLA FIRST) data. The host is also not detected in Galaxy Evolution Explorer (GALEX) UV data, but we obtain a 3-sigma upper limit on the magnitude of $NUV > 20.57$ using co-added data totaling roughly 200s of exposure time. We use this limit when fitting the host galaxy SED in \S\ref{sec:phot}.

Examining data from the ROSAT All-Sky Survey \citep{voges99}, we do not detect the host galaxy to a 3-sigma upper limit of $2.0\times10^{-2}$ counts~s$^{-1}$ in the $0.3-10.0$~keV band. This corresponds to a limit of $1.2\times10^{-13}$~ergs~s$^{-1}$~cm$^{-2}$ ($L_X<6.7\times10^{41}$~~ergs~s$^{-1}$), which provides evidence that the galaxy does not host a strong AGN. Detections of the host in Wide-field Infrared Survey Explorer \citep[WISE;][]{wright10} data corroborate this picture, as the galaxy has a mid-IR (MIR) color of $(W1-W2)\simeq0.06\pm0.06$, implying that any AGN activity in the galaxy is not strong \citep[e.g.,][]{assef13}.

Using the code for Fitting and Assessment of Synthetic Templates \citep[FAST v1.0;][]{kriek09}, we fit stellar population synthesis (SPS) models to the archival GALEX $NUV$ limit, ASAS-SN $V$-band measurement, and 2MASS $JHK_s$ magnitudes of the host galaxy. The fit was done assuming a \citet{cardelli88} extinction law with $R_V=3.1$ and a Galactic extinction of $A_V=0.19$~mag based on \citet{schlafly11}, an exponentially declining star-formation history, a Salpeter IMF, and the \citet{bruzual03} models. We obtained a good SPS fit (reduced $\chi_{\nu}^{2}=1.39$), with the following parameters: $M_{*}=(1.1_{-0.3}^{+0.4})\times10^{10}$~{\msun}, age~$=2.8_{-1.8}^{+1.8}$~Gyr, and a $1\sigma$ upper limit on the star formation rate of $\textrm{SFR}\leq0.002$~{\msun}~yr$^{-1}$. Scaling the stellar mass of {\galname} using the average stellar-mass-to-bulge-mass ratio from the hosts of ASASSN-14ae and ASASSN-14li \citep{holoien14b,holoien16a} implies a bulge mass of $M_B\simeq10^{9.8}$~{\msun}. Using the $M_B-M_{BH}$ relation from \citet{mcconnell13}, we obtain a black hole mass of $M_{BH}=10^{7.1}$~{\msun}, which is significantly more massive than those of the hosts of the previous ASAS-SN TDEs. Fits to the transient spectral energy distribution give no indication of additional extinction related to the host galaxy. In the analyses of the event's SED which follow, we only correct for Galactic extinction.

As there were no archival data in many of our follow-up filters, we lack pre-outburst magnitudes of the host. In order to obtain host magnitudes to use for host flux subtraction, we produced synthesized {\swift} UVOT and Bessel $BVI$ host magnitudes using the best-fit FAST SED. We used a bootstrapping scheme to assess the uncertainties associated with our determination of the host magnitudes. We generated 1000 mock input SEDs based on the observed fluxes and assuming Gaussian errors, and constructed distributions of the magnitudes synthesized from this ensemble. Our adopted host magnitudes fall within $\lesssim 0.2$ mag of the peak of the bootstrapped distributions. These estimated host magnitudes are listed in Table~\ref{table:host_mags2}.

\begin{table}
\centering
\caption{Estimated Photometry of the Host Galaxy}
\label{table:host_mags2}
\begin{tabular}{@{}ccc}
\hline
Filter & Magnitude & Magnitude Uncertainty \\
\hline
$UVW2$ & 21.54 & 0.13 \\
$UVM2$ & 21.20 & 0.17 \\
$UVW1$ & 20.02 & 0.10 \\
$U$ & 18.44 & 0.07 \\
$B_{UVOT}$ & 18.18 & 0.06 \\
$B_{LCOGT}$ & 18.16 & 0.06 \\
$V_{UVOT}$ & 17.28 & 0.05 \\
$V_{LCOGT}$ & 17.20 & 0.05 \\
$I$ & 16.05 & 0.04 \\
\hline
\end{tabular}

\medskip
\raggedright
\noindent 
Estimated 5\farcs{0}-radius aperture magnitudes of the host galaxy synthesized from the host galaxy SED fit.
\end{table}


\subsection{New Photometric Observations}
\label{sec:phot}


\begin{figure*}
\begin{minipage}{\textwidth}
\centering
\subfloat{{\includegraphics[width=0.8\textwidth]{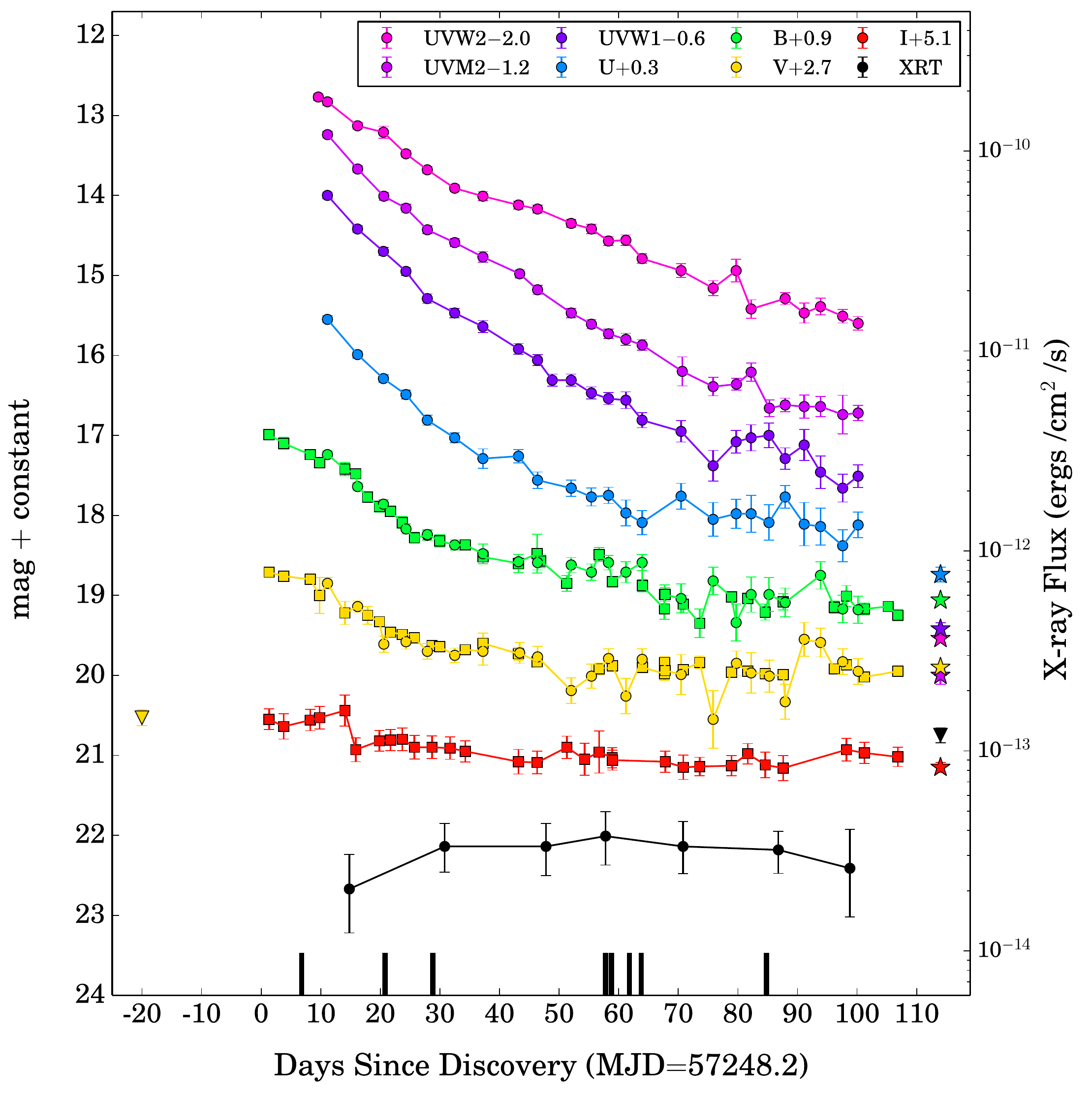}}}
\caption{Light curves of {\name} beginning on the epoch of discovery (MJD$=57248.2$) and spanning 107 days. Follow-up data obtained from {\swift} (X-ray, UV, and optical) are shown as circles and follow-up data obtained by the LCOGT 1-m telescopes (optical) are shown as squares. All optical and UV magnitudes are on the Vega system (left scale), and X-ray fluxes are shown in ergs~cm$^{-2}$~s$^{-1}$ (right scale). The scales are chosen so that time variability has the same meaning for both the X-ray and optical/UV data. The X-ray data points are averages over multiple epochs. The Swift $B$- and $V$-band data were converted to the Johnson-Cousins system using the color corrections found at \protect\url{http://heasarc.gsfc.nasa.gov/docs/heasarc/caldb/swift/docs/uvot/}. The data are not corrected for extinction and error bars are shown for all points, but can be smaller than the data points. The previous $V$-band non-detection from ASAS-SN is plotted as a triangle at $-20$ days, the date of our previous observation of the host galaxy. Estimated 5\farcs0 aperture magnitudes measured from the host galaxy SED fit (stars) and an X-ray upper limit from RASS (triangle) are shown at $+114$ days. Vertical bars at the bottom of the figure indicate the dates of spectroscopic observations. Although the rapid rise starting after our previous non-detection at $-20$ days was missed, the data still show that {\name} brightened considerably in the UV and in the bluer optical filters, with the largest increase being roughly 3.4 magnitudes in the {\swift} $UVM2$ band. Table~\ref{tab:phot} contains all the follow-up photometric data.}
\label{fig:lightcurve}
\end{minipage}
\end{figure*}

After the transient was classified as a TDE, we obtained a series of 26 {\swift} XRT and UVOT target-of-opportunity (ToO) observations. The UVOT observations were obtained in 6 filters: $V$ (5468~\AA), $B$ (4392~\AA), $U$ (3465~\AA), $UVW1$ (2600~\AA), $UVM2$ (2246~\AA), and $UVW2$ (1928~\AA) \citep{poole08}. We extracted source counts from a 5\farcs0 radius region and sky counts from a $\sim$40\farcs0 radius region using the UVOT software task {\sc uvotsource} and converted these count rates into magnitudes and fluxes using the most recent UVOT calibrations \citep{poole08,breeveld10}. 

The XRT operated in Photon Counting mode \citep{hill04} for our observations. The data from all epochs were reduced with the software task {\sc xrtpipeline}. We extracted X-ray source counts and background counts with {\sc xrtselect} using a region with a radius of 10 pixels (23\farcs{6})  centered on the source position and a source-free region with a radius of 100 pixels (235\farcs{7}), respectively. While the source is undetected in most individual epochs, we detect X-ray emission consistent with the position of the transient after combining the signal from multiple exposures. We assume a blackbody plus power law model (see \S\ref{sec:sedanal}) and Galactic \ion{H}{i} column density of 5.6$\times 10^{20}$ cm$^{-2}$ \citep{kalberla05} to convert the detected counts into fluxes, with a count to flux conversion factor of $5.2\times 10^{-11}$ ergs s$^{-1}$ (cm$^2$ counts s$^{-1})^{-1}$ for the absorbed spectrum and $8.5\times 10^{-11}$ ergs s$^{-1}$ (cm$^2$ counts s$^{-1})^{-1}$ for the absorption-corrected spectrum.

We also obtained $BVI$ images with the Las Cumbres Observatory Global Telescope Network \citep[LCOGT;][]{brown13} 1-m telescopes at Siding Spring, South African Astronomical, and Cerro Tololo Inter-America Observatories. We measured magnitudes using aperture photometry with a 5\farcs0 aperture radius in order to match the host galaxy and {\swift} measurements. Photometric zero-points were determined using several stars in the field.

In order to constrain any offset between the source of the outburst and the nucleus of the host galaxy, we astrometrically aligned an $I$-band image of the transient taken on 2015 September 4 with the LCOGT 1-m telescope at Sutherland, South Africa with the archival DSS image of the host galaxy. We then measured the offset between the centroid position of the transient in the LCOGT image and that of the host galaxy in the DSS image, finding an offset of $0.24\pm0.05$~arcseconds ($254.8\pm53.1$~parsecs). We performed the same analysis using two LCOGT $V$-band images from 2015 September 4 and 2015 November 23, after the transient had faded. In order to increase the accuracy of the transient position, we first used \textsc{Hotpants}\footnote{\url{http://www.astro.washington.edu/users/becker/v2.0/hotpants.html}, an implementation of the \citet{alard00} algorithm for image subtraction, to subtract the 2015 November 23 template image from the 2015 September 4 image to isolate the transient flux. The centroid position of the transient in the subtracted image is located 0\farcs{08} West and 0\farcs{05} North of the centroid position of the host in the template image, giving a total offset of $0.09\pm0.06$~arcseconds ($96.3\pm62.8$~parsecs).} Figure~\ref{fig:finding_chart} shows the two LCOGT $V$-band images used to measure the offset. Higher resolution data, particularly in the UV, could improve these estimates.

Figure~\ref{fig:lightcurve} shows the X-ray, UV, and optical light curves of {\name}. Table~\ref{tab:phot} and Table~\ref{tab:xray} give the UVOT/$VRI$ magnitudes and XRT flux measurements, respectively. The photometric observations cover the period from MJD 57248.2 (the epoch of discovery) through our latest epoch of observations on MJD 57355.0, a span of 106.8 days. The data shown in Figure~\ref{fig:lightcurve} are presented without extinction correction or host flux subtraction. Also shown in Figure~\ref{fig:lightcurve} are the host magnitudes estimated from the host SED fit. Our observations show that {\name} brightened considerably with respect to the host galaxy in the UV and bluer optical filters in the short period between our previous observation on 2015 July 25 and our detection of the transient on 2015 August 14. The largest increase came in the {\swift} $UVM2$ band, where it brightened by $\Delta m_{UVW2}\sim-6.8$, while the $V$-band increase was weaker, at $\Delta m_{V}\sim-1.2$. This large UV increase is additional evidence against typical AGN variability, which is normally of much smaller magnitude \citep[e.g.,][]{gezari12b,macleod12}. 


\subsection{New Spectroscopic Observations}
\label{sec:spec}

\begin{figure}
\centering
\includegraphics[width=0.48\textwidth]{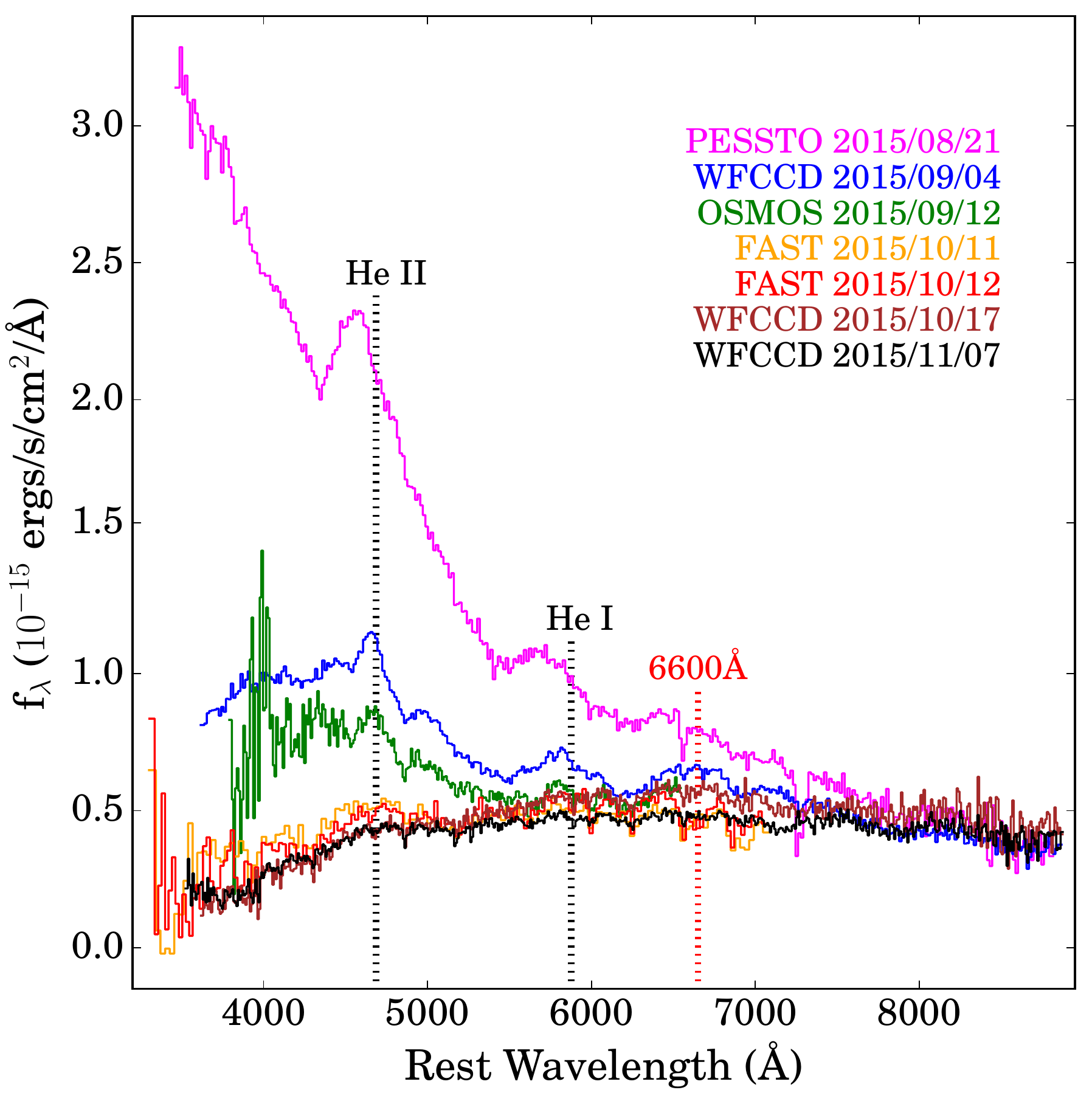}
\caption{Spectral time-sequence for {\name}. The dates of the observations and the instruments used are listed in matching colors and order to the spectra. (See Table~\ref{tab:spectra} for more information.) The prominent \ion{He}{i}~5876{\AA} and \ion{He}{ii}~4686{\AA} emission features are indicated by black vertical dotted lines, while the feature near $6600${\AA} which appears in the 2015 September 4 spectrum (see \S\ref{sec:specanal}) is indicated by a vertical red dotted line. The transient spectra show many broad emission features and blue continuum emission in early epochs which rapidly fade, with later spectra resembling that of the host galaxy rather than that of the transient.}
\label{fig:spectra}
\end{figure}

We obtained spectra of {\name} spanning 69 days between UT 2015 September 04 and UT 2015 November 07. The spectrographs used for these observations were the Ohio State Multi-Object Spectrograph \citep[OSMOS;][]{martini11} mounted on the MDM Observatory Hiltner 2.4-m telescope ($4200-6800$~\AA, $\rm R\sim 4$~\AA), the Fast Spectrograph \citep[FAST;][]{fabricant98} mounted on the Fred L. Whipple Observatory Tillinghast 1.5-m telescope ($3700-9000$~\AA, $\rm R\sim 3$~\AA), and the Wide Field Reimaging CCD Camera (WFCCD) mounted on the Las Campanas Observatory du Pont 2.5-m telescope ($3700-9500$~\AA, $\rm R\sim 7$~\AA). The spectra were reduced using standard techniques in IRAF, and we applied telluric corrections using observations of spectrophotometric standard stars from the same nights. Each spectrum was scaled to match the $V$-band photometry. Figure~\ref{fig:spectra} shows a time-sequence of the flux-calibrated follow-up spectra as well as the initial spectrum obtained by PESSTO \citep{asassn15oi_spec_atel}. Summary information for each spectrum is listed in Table~\ref{tab:spectra}.


The key characteristics of the early spectra of {\name} are a strong blue continuum, consistent with the photometric measurements, and the presence of broad helium lines in emission. The emission features are broad and asymmetric, with initial widths of $\sim10000-20000$~km~s$^{-1}$ that narrow over time. In the initial spectrum obtained by PESSTO, the helium lines are blueshifted by approximately $6000-8000$~km~s$^{-1}$, but in later spectra little-to-no apparent shift relative to the systemic velocity is observed. The transient features fade rapidly, with spectra later than 2015 September 12 showing no blue continuum and absorption features consistent with the transient's host galaxy. Unlike other recently discovered TDEs \citep[e.g.][]{arcavi14,holoien16a,french16}, the later host-dominated spectra do not show features, such as H$\delta$ absorption with a large equivalent width, that are consistent with the host being a post-starburst galaxy. We further analyze the features of these spectra and compare them to other TDEs and supernovae in \S\ref{sec:analysis}.


\section{Analysis}
\label{sec:analysis}


\subsection{SED Analysis}
\label{sec:sedanal}

\begin{figure}
\centering
\includegraphics[width=0.48\textwidth]{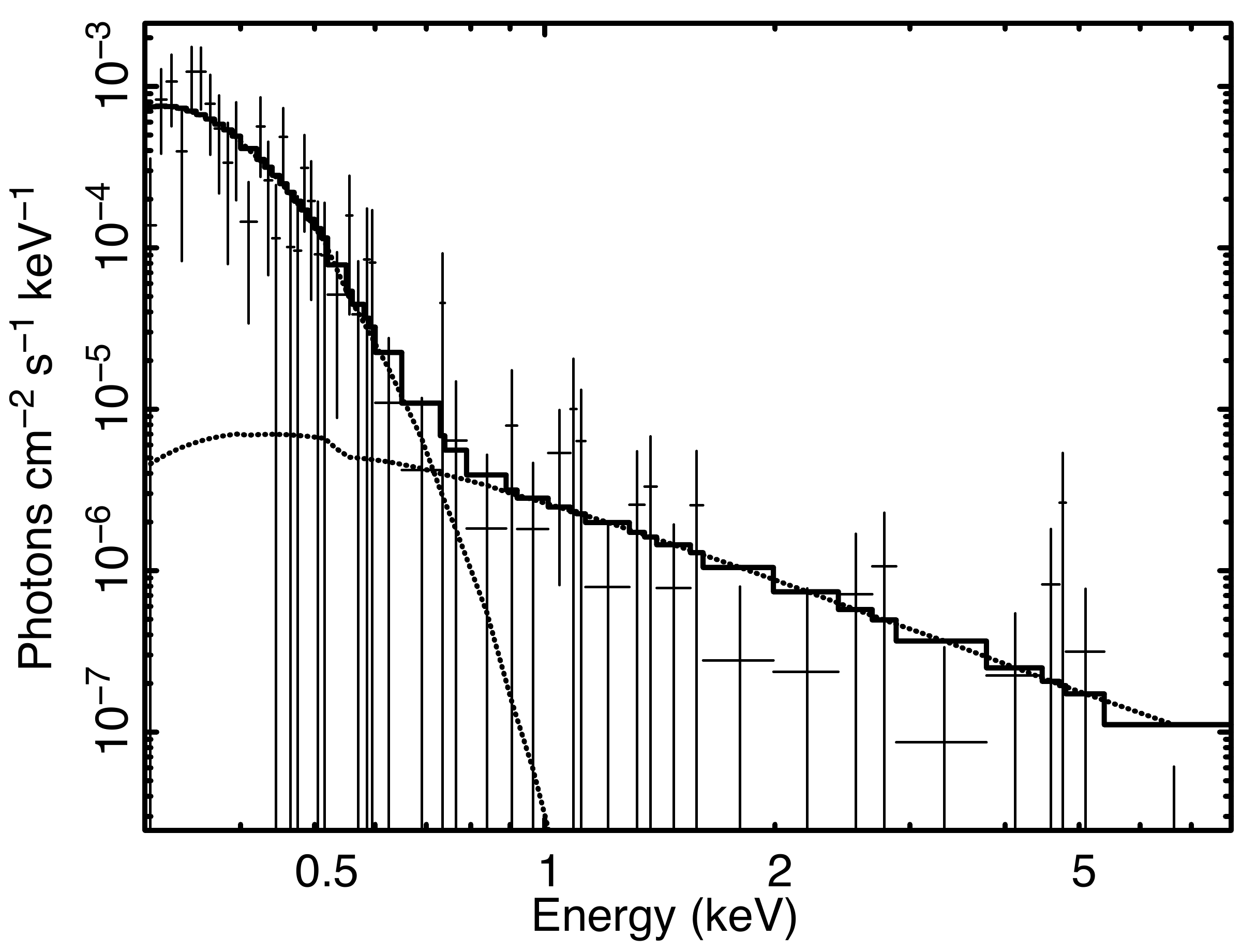}
\caption{Unfolded X-ray spectrum (unbinned) of ASASSN 15oi fitted with an absorbed black body plus power law model, with a blackbody temperature of 49$^{+10}_{-9}$ eV and a power law index of $\Gamma=1.76^{+1.50}_{-0.84}$. The dotted lines display the black body and power law models and the solid line shows the combined model. The data were fitted with Cash Statistics \citep{cash79}.}
\label{fig:xray_spec}
\end{figure}

Combining all 61.4 ks of observations, we were able to extract an X-ray spectrum of the source. The spectrum was analyzed with {\sc xspec} 12.8.2 \citep{arnaud96} using Cash statistics \citep{cash79}. A fit with a single power law model does not result in an acceptable fit. The X-ray spectral slope is very steep ($\Gamma$=5.8) with large residuals at energies above 2.0 keV. We found an acceptable fit using an absorbed blackbody plus power law model with the absorption column density fixed to the Galactic value. This results in a blackbody temperature at an energy of 49$^{+10}_{-9}$ eV and a photon index for the hard energy component of $\Gamma=1.8^{+1.5}_{-0.8}$. The absorbed and unabsorbed fluxes in the 0.3-10 keV band are 8.0$^{+0.4}_{-0.3}\times 10^{-14}$ ergs s$^{-1}$ cm$^{-2}$ and 18.7$^{+0.9}_{-0.6}\times 10^{-14}$ ergs s$^{-1}$ cm$^{-2}$, respectively. The X-ray spectrum and the model fits are shown in Figure~\ref{fig:xray_spec}.


There are two challenges to interpreting the X-ray detections. First, the archival ROSAT data are not deep enough to rule out a pre-existing source with the presently observed X-ray luminosity ($L_X< 6.7\times10^{41}$ versus $L_X = 4.8\times10^{41}$~ergs/s). Second, the X-ray emission we observe during the transient is consistent with constant luminosity. We can rule out X-ray emission dominated by sources other than an AGN. Low Mass X-ray Binaries (LMXBs) in old stellar populations, like the FAST models for the SED of {\galname}, produce X-rays. Based on the correlations of \citet{kim04} between the $K$-band luminosities of galaxies and the integrated X-ray emission by their LMXBs, however, this contribution should only be of order $10^{40}$~ergs~s$^{-1}$, well below the observed luminosity. Thus, the X-ray emission is almost certainly dominated by AGN activity, and seems likely to be associated with the present transient, but this association is not required by the available data. Late-time observations of the X-ray evolution should illuminate the source of this emission.

\begin{figure}
\centering
\subfloat{{\includegraphics[width=0.48\textwidth]{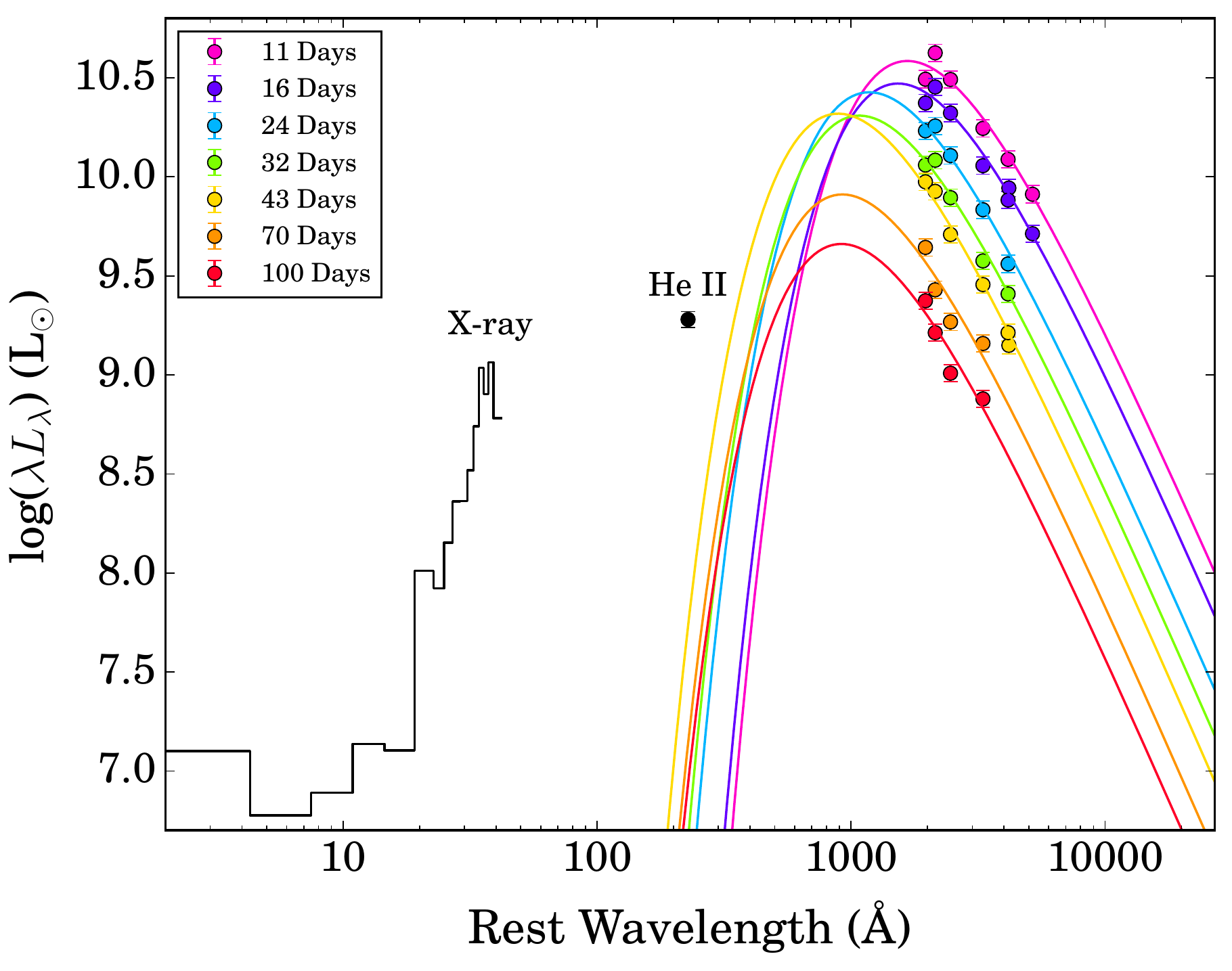}}}
\caption{Evolution of the SED of {\name} shown in different colors (points) and the corresponding best-fitting blackbody models for each epoch (lines). All data have been corrected for Galactic extinction and host contribution. Error bars are shown but can be smaller than the data points. All fits shown were made assuming the temperature prior described in \S\ref{sec:sedanal}. The X-ray spectral model inferred from the average of all the X-ray data (see Figure~\ref{fig:xray_spec} and the ionizing luminosity implied by the \ion{He}{ii} 4686\AA~line (see \S\ref{sec:specanal}) are shown in black. Neither the X-ray emission nor the ionizing luminosity for the \ion{He}{ii} line can be explained by the single blackbody model inferred from the optical/UV data, and a more complex model is likely needed.}
\label{fig:sed_evol}
\end{figure}

Figure~\ref{fig:sed_evol} shows the evolution of the UV-optical SED of {\name}. Using the 5\farcs0 aperture host magnitudes estimated from the host SED fit, we produced host- and extinction-corrected light curves for all optical and UV filters. We then fit these host-subtracted fluxes with blackbody models using Markov Chain Monte Carlo (MCMC) methods, as was done for the previous ASAS-SN TDEs \citep{holoien14b,holoien16a}. Due to there being uncertainty in the actual host magnitude, we only included epochs for which the transient's flux was greater than 50\% of the host galaxy flux when performing these SED fits. This criteria only excludes later epochs of observations in optical filters, as the UV emission remains well above our estimated host levels throughout our observations.

Unlike the previous ASAS-SN TDEs, the data for {\name} indicate that the source's temperature increases before (probably) leveling off in later epochs. Because of this, we fit the temperature using a changing prior based on initial, unconstrained fits. For epochs within 10 days of discovery, the temperature was fit with a prior of $T=2\times10^4$~K. For later epochs, we fit the data with a prior of $T=(2+(\Delta t-10)/2)\times10^4$~K, where $\Delta t=MJD-57248.2$ and the temperature is capped at $T=40000$~K, as this provides a better fit than allowing the temperature to increase arbitrarily and is consistent with the color evolution (see \S\ref{sec:tde_comp}). These priors were applied with a $\log$ uncertainty of $\pm0.05$ dex in all epochs. Due to the fact that our {\swift} data do not span the peak of the SED for the later, hotter epochs, the temperatures in these phases are not well-constrained, although the prior used for these fits does capture the steepening spectral slope. Also shown in Figure~\ref{fig:sed_evol} are the X-ray spectral model (see Figure~\ref{fig:xray_spec}) and the ionizing luminosity implied by the \ion{He}{ii} 4686\AA~line luminosity (see \S\ref{sec:specanal}). It can be clearly seen that the blackbody model inferred from the optical/UV data is not enough to explain the observed X-ray and line emission.


\begin{figure}
\centering
\subfloat{{\includegraphics[width=0.95\linewidth]{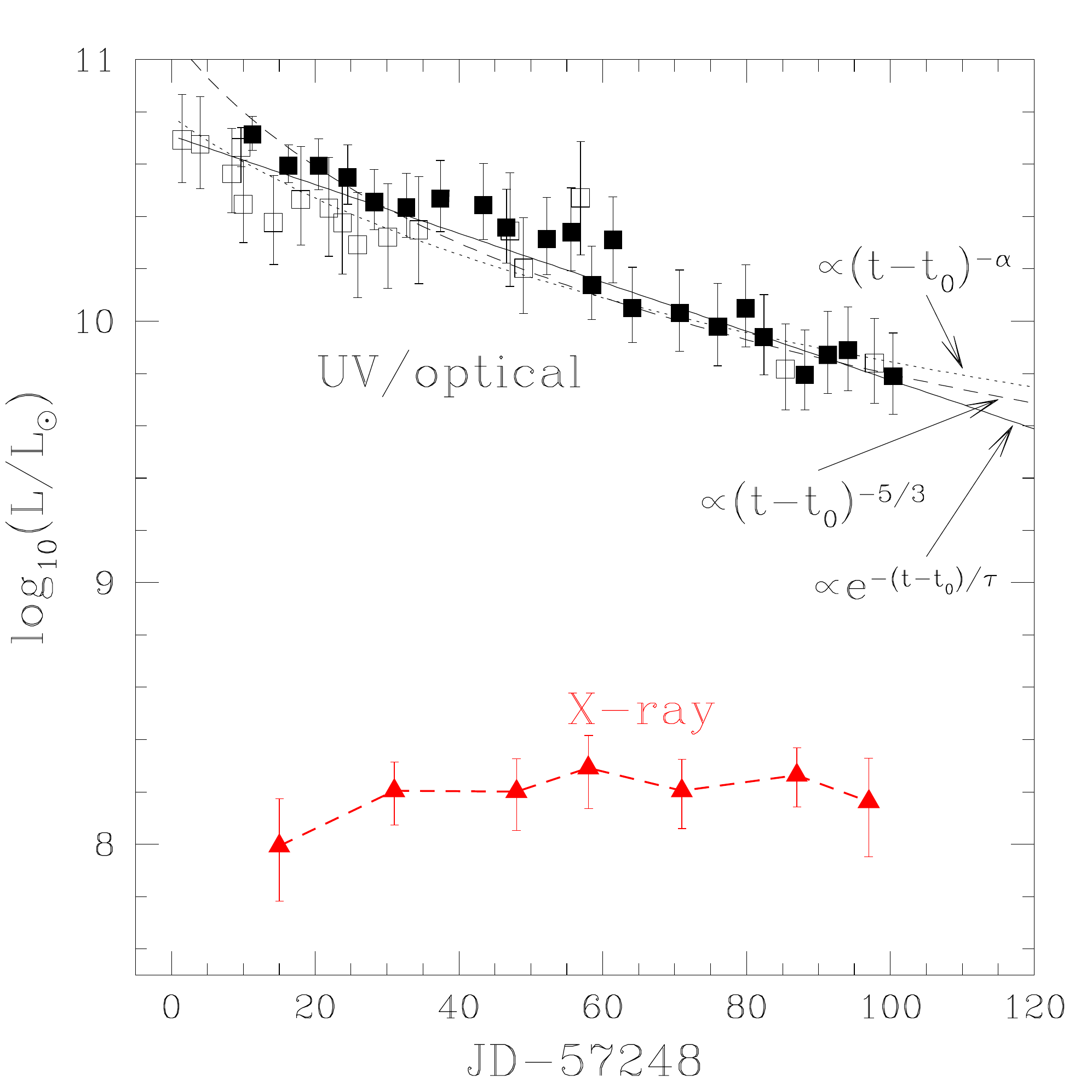}}}
\caption{Evolution of \name's X-ray (red triangles) and optical/UV (black squares) luminosity over time assuming the emission is consistent with a blackbody with the temperature prior described in the text. Filled points indicate those epochs where 4 or more data points were used to fit the temperature, and unfilled points indicate those epochs where fewer data points were used. The dashed line shows the best $L\propto (t-t_0)^{-5/3}$ power law fit with $t_0=57228.0$, while the dotted line shows the best unconstrained power law fit $L\propto (t-t_0)^{-\alpha}$ with $t_0=57212.3$ and the power law index $\alpha=1.62$. The solid line shows the best exponential fit, $L\propto e^{-(t-t_0)/\tau}$, with $t_0=57241.6$ and $\tau=46.5$~days. While all three models fit the data reasonably well, the exponential is the best match to the data.}
\label{fig:lum_evol}
\end{figure}

The optical/UV luminosity of {\name}, shown in Figure~\ref{fig:lum_evol}, fades steadily over the $\sim3$ months after the initial discovery with the temperature constrained as described above. As was the case with previous ASAS-SN TDEs, the luminosity evolution inferred from the blackbody fit at early times can be well-described by an exponential $L= L_0e^{-(t-t_0)/\tau}$ \citep{holoien14b,holoien16a}, with best-fit parameters of $\log{L_0/{\lsun}}=10.8$ and $\tau\simeq46.5$~days for $t_0=57241.6$. This decay rate is quite similar to that observed for ASASSN-14ae ($\tau \simeq39$~days for $t_0=56682.5$). However, the late-time light curves of ASASSN-14ae and ASASSN-14li can also be fit by power-law profiles \citep[][Brown et al. 2016b \emph{in prep.}]{brown16}. For this reason, we also fit the luminosity evolution of {\name} with the popular $L=L_0 (t-t_0)^{-5/3}$ power law, finding a best fit of $\log{L_0/{\lsun}}=13.3$ and $t_0=57228.0$, although this is a significantly worse fit than the exponential ($\chi^2=55.4$ vs. $\chi^2=15.1$). For the $t^{-5/3}$ power law model, we used the constraint $57228 \leq t_0 \leq 57248$ so that the disruption begins sometime between our previous non-detection and the discovery of the transient. If we relax this constraint and allow $t_0$ and the power law index to vary freely, we find a best fit of $\log{L_0/{\lsun}}=13.3$ and $t_0=57212.3$ with a power law index of $\alpha=1.62$. This model is still a poorer fit than the exponential model ($\chi^2=19.0$ vs. $\chi^2=15.1$), and it would require a fairly rapid rise to be consistent with our flux limits from numerous epochs of data obtained between MJD 57212 and MJD 57228. With a longer temporal baseline, we will be better able to distinguish between exponential and power-law decline models \citep[e.g.,][]{brown16}.

This temperature and luminosity evolution is inconsistent with what would be expected if {\name} were a supernova, which typically exhibit rapidly declining temperatures along with constant or declining luminosity \citep[e.g.,][]{miller09,botticella10,inserra13,graham14}. Normal AGN activity is ruled out by the archival host photometry, leaving a TDE as the most likely explanation for ASASSN-15oi.

In contrast to the optical/UV luminosity evolution, the X-ray luminosity remains constant for the duration of the observations. However, even in later epochs, the X-ray luminosity is nearly two orders of magnitude lower than the optical/UV luminosity, and even if it is associated with the TDE, it has little impact on the overall energy output of the source. As shown in Figure~\ref{fig:sed_evol}, a single blackbody model is incapable of fitting both the X-ray and optical/UV emission. Thus, if the X-ray emission is associated with the TDE, we infer that the X-ray emission likely comes from a different region of the TDE than the optical/UV emission. Including the UV luminosity implied by the \ion{He}{ii} emission, ASASSN-15oi is another piece of evidence pointing to the need for more complex models of TDE emission \citep[e.g.,][]{metzger15,roth15}.

Integrating the combined optical/UV and X-ray luminosity curves gives a total radiated energy of $\sim6.6\times10^{50}$~ergs for the $\sim3.5$ months of follow-up observations. The uncertainties in this estimate are predominantly systematic due to the poorly constrained temperatures at later times.


\subsection{Spectroscopic Analysis}
\label{sec:specanal}

Though the SED evolution of {\name} is largely consistent with the transient being a tidal disruption flare, the rapid evolution and absolute magnitude are unique among TDEs discovered at optical wavelengths, and it is worth examining other possible sources of the observed emission. The most likely alternative explanation for the transient is that it was a supernova, as archival X-ray, UV, infrared, and radio observations do not show any indication that {\galname} hosts a typical AGN. While supernovae evolve on the rapid timescales observed for {\name}, a spectroscopic comparison between {\name} and supernovae reveals that this is not the case.


\begin{figure}
\centering
\includegraphics[width=0.48\textwidth]{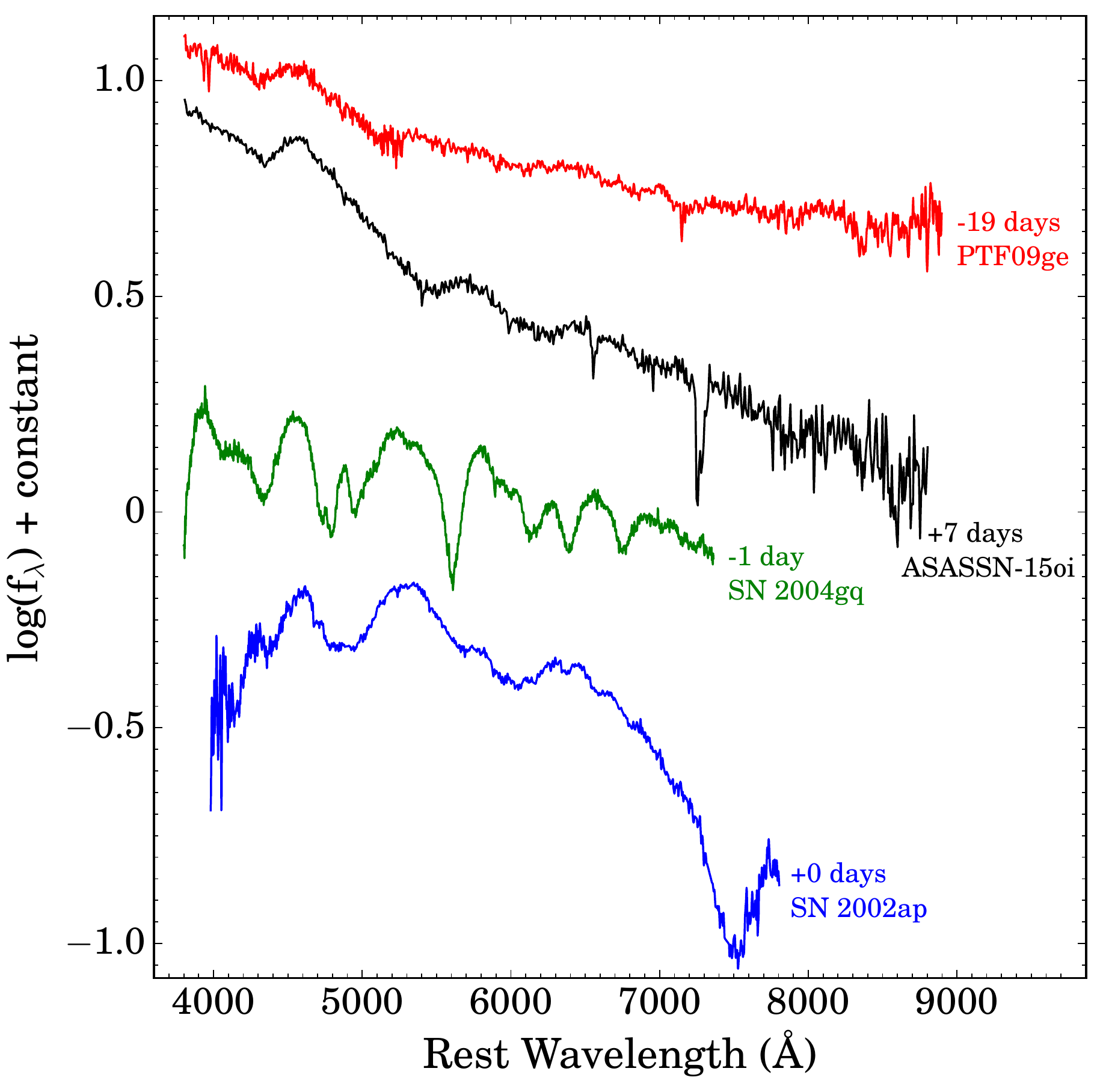}
\caption{Comparison between spectra taken at or before maximum of {\name} \citep{asassn15oi_spec_atel}, the He-rich TDE PTF09ge \citep{arcavi14}, the Type Ib supernova SN 2004gq \citep{modjaz14}, and the broad-line Type Ic supernova SN 2002ap \citep{galyam02}. The epoch of each spectrum in days-since maximum-light is given next to each spectrum. (The epoch of {\name} is days-since-discovery, since the peak date is not known.) Spectroscopically, {\name} is far more similar to PTF09ge than it is to either supernova, which have different emission and absorption features and different continuum shapes.} 
\label{fig:spec_comp}
\end{figure}

{\name} shows broad He emission lines in its early spectra but little-to-no H emission. If {\name} were a supernova, it would likely be a Type Ib or Type Ic-bl: the former because SNe Ib show He emission with no H emission, and the latter because broad-line SNe Ic exhibit line widths comparable to those observed in {\name}. Figure~\ref{fig:spec_comp} shows a comparison between spectra taken at or before maximum for {\name} \citep{asassn15oi_spec_atel}, the He-rich TDE PTF09ge \citep{arcavi14}, the Type Ib supernova SN 2004gq \citep{modjaz14}, and the broad-line Type Ic supernova SN 2002ap \citep{galyam02}. While SN 2004gq does show the helium emission lines present in the spectrum of {\name}, there are also numerous absorption features that are not present in the spectrum of {\name}. Similarly, SN 2002ap shows broad emission features, but does not have the same helium emission lines, and the shape of the continuum is very different. In contrast, the spectrum of PTF09ge is very similar to that of {\name}, as both objects show a strong blue continuum, a broad \ion{He}{ii} 4686\AA~emission feature, and no hydrogen emission. Given this comparison, we conclude that it is very unlikely that {\name} is a supernova.

\citet{arcavi14} noted that optically-selected TDEs span a continuum from H-rich to He-rich spectroscopic features. As indicated in Figure~\ref{fig:spec_comp}, early follow-up spectra of {\name} are highly consistent with those of He-rich TDEs like PS1-10jh and PTF09ge \citep{gezari12b,arcavi14}. The prominent features of the earliest spectra are a strong blue continuum with broad \ion{He}{ii} and \ion{He}{i} emission features. The earliest spectrum, obtained by PESSTO on 2015 August 21, shows \ion{He}{ii} 4686\AA~and \ion{He}{i} 5876\AA~emission features with $\rm{FWHM}_{\ion{He}{ii}}\simeq 19700$~km~s$^{-1}$ and $\rm{FWHM}_{\ion{He}{i}}\simeq 18800$~km~s$^{-1}$, respectively, that are blueshifted by $\sim 6000-8000$~km~s$^{-1}$. The second spectrum, obtained with the du~Pont+WFCCD on 2015 September 4, shows some interesting changes. The \ion{He}{ii} 4686\AA~feature is still prominent but has a smaller width ($\rm{FWHM}_{\ion{He}{ii}}\simeq 9200$~km~s$^{-1}$) compared to the first spectrum and the peak of the line is only blueshifted by $\sim 1400$~km~s$^{-1}$. Conversely, the \ion{He}{i} 5876\AA~line has remained broad, with $\rm{FWHM}_{\ion{He}{i}}\simeq 18800$~km~s$^{-1}$. There are two new broad bumps ($\rm{FWHM}\sim 10000$~km~s$^{-1}$) that appear around the strong \ion{He}{ii} 4686\AA~line with peak wavelengths at $\sim 4435$\AA~and $\sim 5000$\AA. These features are consistent with the \ion{He}{i} 4472\AA~and \ion{He}{i} 5016\AA~transitions, blueshifted by $1000-2500$~km~s$^{-1}$. This spectrum also shows a new feature at $\sim 6600$\AA~which exhibits an integrated flux that is comparable to the \ion{He}{ii} 4686\AA~line flux, indicated with a red dotted line in Figure~\ref{fig:spectra}. This is near the position of H$\alpha$, but it is likely that this feature is due at least in part to the \ion{He}{ii} 6560\AA~$n=6\rightarrow4$ line, which is expected to be an order of magnitude weaker than \ion{He}{ii} 4686\AA~in photoionized gas \citep{osterbrock89,gezari15}.

Unlike what has been seen in previous TDEs, the spectroscopic features of {\name} evolve rapidly. Between the classification spectrum obtained on UT 2015 August 21 and the first follow-up spectrum obtained on UT 2015 September 4, the blue continuum in the spectra of {\name} fades and the \ion{He}{ii} line narrows considerably. The emission lines seem to have completely faded by the time of the latest follow-up spectrum obtained on UT 2015 November 7, slightly less than 3 months after discovery.

After correcting for Galactic reddening and subtracting an estimated continuum from the PESSTO classification spectrum, we estimate the luminosity of the \ion{He}{ii} 4686\AA~line to be $L_{\ion{He}{ii}}\simeq(1.1\pm0.1)\times10^{42}$~ergs~s$^{-1}$, where the dominant source of uncertainty is in setting the continuum level. In the UT 2015 September 4 spectrum, the line has faded to a luminosity of $L_{\ion{He}{ii}}\simeq(1.7\pm0.5)\times10^{41}$~ergs~s$^{-1}$. We are unable to obtain estimates of the line luminosity in later spectra, as the line has faded and the continuum is too uncertain to provide an accurate subtraction. Assuming case B recombination \citep{osterbrock89}, the line luminosity from the classification spectrum implies an ionizing luminosity of $L_i\simeq(7.3\pm0.7)\times10^{42}$~ergs~s$^{-1}$, while the line luminosity from the September 4 spectrum implies an ionizing luminosity of $L_i\simeq(1.1\pm0.3)\times10^{42}$~ergs~s$^{-1}$. While this is a rapid drop, in both cases the line luminosity is greater than the peak line luminosity measured for ASASSN-14li \citep{holoien16a}. As shown in Figure~\ref{fig:sed_evol}, the optical/UV emission is unable to provide the requisite ionizing luminosity needed to power the observed \ion{He}{ii} line, and additional ionizing flux must be coming from a different region of the TDE.

It is not uncommon for the emission features in TDE spectra to narrow and become less luminous over time \citep[e.g.,][]{holoien14b,holoien16a}, which is the opposite behavior to that observed in reverberation mapping studies of quasars, where the line width broadens as the luminosity decreases \citep[e.g.][]{peterson04,denney09}. However, while the spectroscopic evolution of {\name} is similar to that of other optically-selected TDEs, it occurs on a significantly faster timescale. This rapid line evolution could be indicating that the ``reprocessing layer'' responsible for the optical/UV emission and emission lines, is becoming optically thin or is no longer in our line-of-sight to the TDE. It is also worth noting that ASASSN-14ae, which had a similar $\sim3$ week gap between the previous epoch of observation and the epoch of discovery, also showed a blueshift in its broad emission features in its early spectra which was not present in later spectroscopic observations \citep{holoien14b}. While ASASSN-14li did not show any blueshift in its spectroscopic observations \citep{holoien16a}, it also had a significantly longer gap ($\sim3$ months) between the epoch of discovery and the previous epoch of observations due to Sun constraints. It is possible that an early blueshift in the emission lines of optically-selected TDEs is common, and that it was not seen in ASASSN-14li due to the gap in observations.

As discussed by \citet{french16}, optically-selected TDEs have shown a preference for unusual ``quiescent Balmer-strong galaxies''---those galaxies whose spectra show little-to-no line emisison but strong Balmer line absorption, indicating a recent period of intense star formation. While archival spectra of {\galname} are not available, our later epochs of follow-up spectra are dominated by the host galaxy, allowing us to determine whether the host of {\name} also falls into this rare class of galaxies. From the 2015 November 7 du Pont/WFCCD spectrum, we measure H$\delta$ and H$\alpha$ absorption features with $\textrm{EW}=1.1$~\AA~and $\textrm{EW}=0.6$~\AA, respectively. In \citet{french16}, a quiescent Balmer-strong galaxy is defined as having absorption with a Lick H$\delta_A$ index of H$\delta_A>4$~\AA~(equivalent to H$\delta~\textrm{EW}>3$~\AA) and H$\alpha~\textrm{EW}<3$~\AA. With much narrower H$\delta$ absorption, and H$\alpha$ in absorption rather than emission, {\galname} clearly does not fall into the same class of galaxy as the TDE hosts discussed in \citet{french16}, and is more consistent with the old stellar population indicated by the FAST SED fits.


\subsection{Comparison of ASAS-SN TDEs}
\label{sec:tde_comp}

ASAS-SN has now found three nearby TDEs, all of which were intensively observed with {\swift} and ground-based telescopes for months after discovery, allowing us to compare their photometric properties and the inferred temperatures and luminosities at early and late epochs. Such an analysis was not possible with many earlier TDEs, due to the fact that they were fainter, and late-time observation in many cases was not possible. For optical TDEs where such analyses were possible, previous studies have shown that they typically show little-to-no color evolution or change in temperature \citep[e.g.,][]{velzen11,gezari12b,arcavi14,chornock14}, and while we find that the first two ASAS-SN TDEs follow this pattern, {\name} appears to be different. All data for ASASSN-14ae and ASASSN-14li shown below are taken from \citet{holoien14b} and \citet{holoien16a}, respectively.


\begin{figure}
\centering
\includegraphics[width=0.48\textwidth]{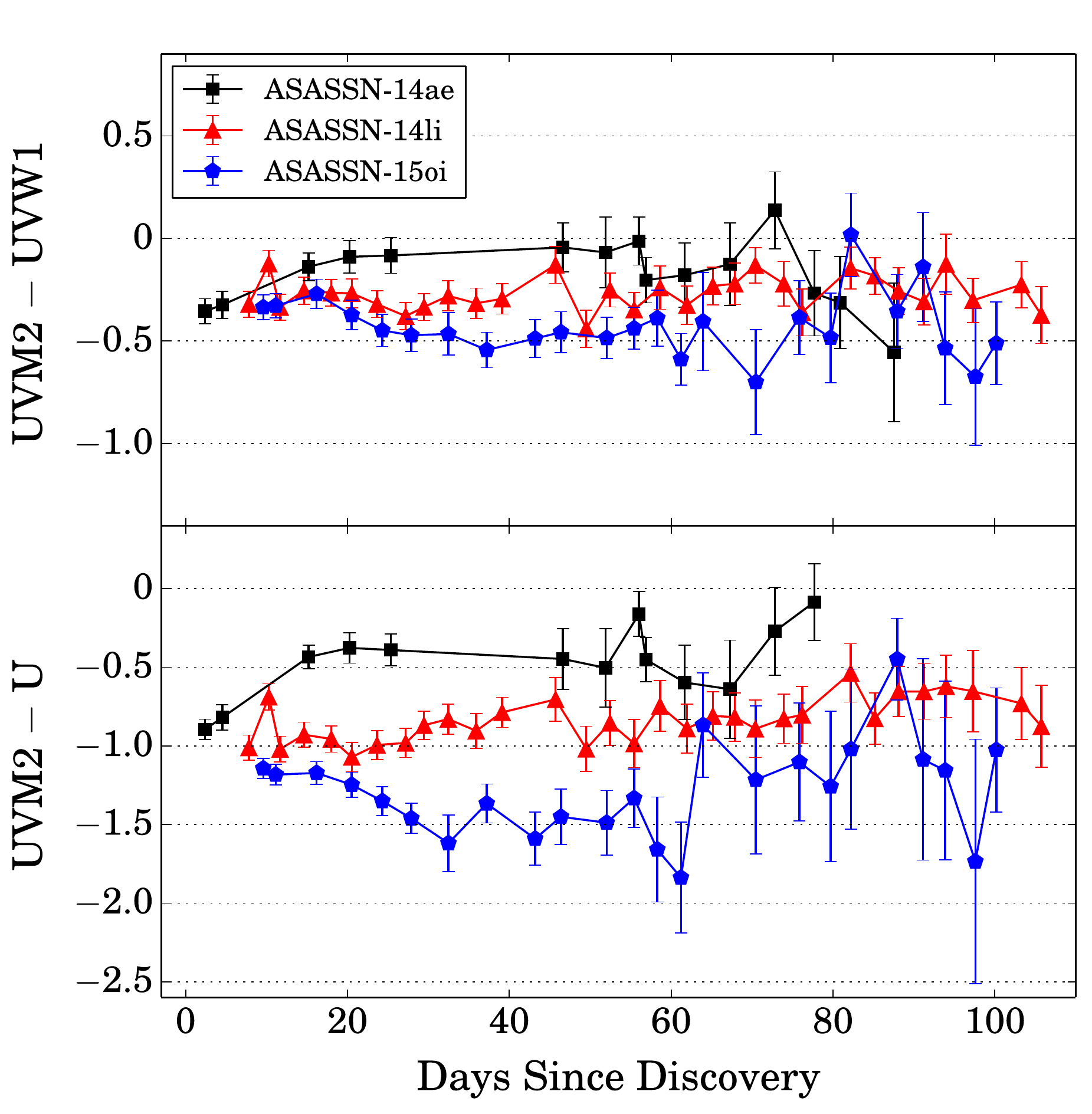}
\caption{Comparison of ($UVM2-UVW1$) (top panel) and ($UVM2-U$) (bottom panel) color evolution between {\name} (blue pentagons); ASASSN-14ae \citep[][black squares]{holoien14b}; and ASASSN-14li \citep[][red triangles]{holoien16a}. Extinction correction and host flux subtraction has been applied to all objects. All three TDEs show little evolution in ($UVM2-UVW1$) and remain quite blue in both colors for months after discovery. {\name} does become bluer in ($UVM2-U$) before leveling off in later epochs, indicating an increasing temperature.}
\label{fig:color_comp}
\end{figure}


\begin{figure}
\centering
\includegraphics[width=0.48\textwidth]{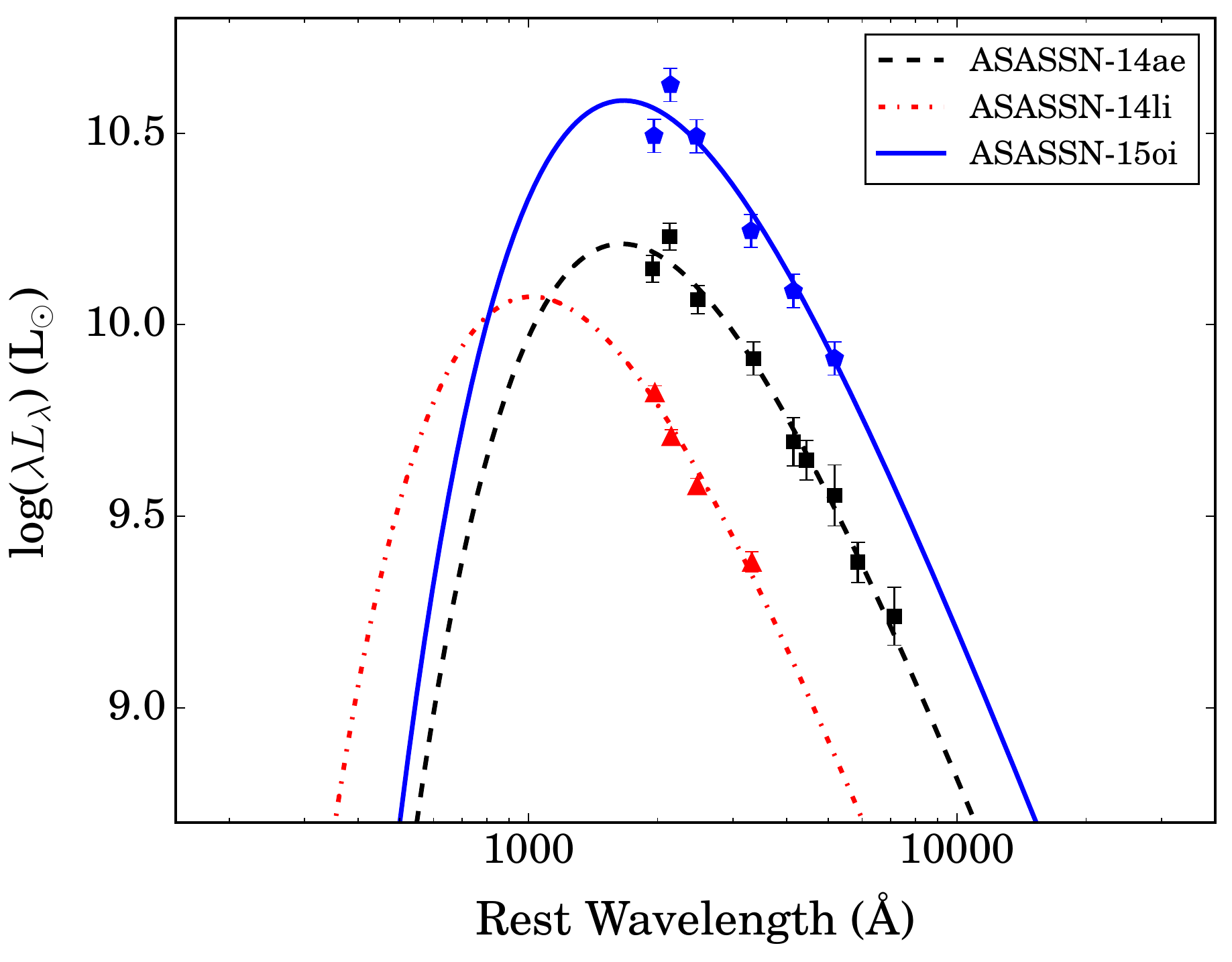}
\caption{Comparison of the blackbody SEDs inferred from early epoch optical/UV data for {\name} (blue solid line, 11 days after discovery), ASASSN-14ae (black dashed line, days after discovery), and ASASSN-14li (red dash-dotted line, 8 days after discovery). In all three cases, the temperature was fit to host-subtracted and extinction-corrected {\swift} and ground photometric data, which are also shown on the figure in the same colors and symbols used in Figure~\ref{fig:color_comp}. ASASSN-14li and {\name} were fit with a temperature prior. Though more luminous, {\name} exhibits a blackbody temperature that is not uncommon for an optically-selected TDE.}
\label{fig:sed_comp}
\end{figure}

From the host-subtracted light curves we produced color evolution curves for {\name}, ASASSN-14ae, and ASASSN-14li. Figure~\ref{fig:color_comp} compares the {\swift} ($UVM2-UVW2$) and ($UVM2-U$) colors of the three TDEs. All three objects show blue UV-UV and UV-optical colors in all epochs, which is expected for optically-selected TDEs. The ($UVM2-UVW2$) colors of all three objects are remarkably similar and show little evolution for all three objects, but the ($UVM2-U$) colors do show some variation: while they remain blue overall, ASASSN-14ae becomes slightly redder and {\name} becomes slightly bluer before leveling off in later epochs, while ASASSN-14li shows little evolution. The blue evolution of the ($UVM2-U$) color for {\name} is in line with the increasing temperature described in \S\ref{sec:sedanal}, and it is not unexpected that {\name} would differ slightly from the other two TDEs, as ASASSN-14ae and ASASSN-14li both had roughly constant temperatures in all epochs. In all cases, the magnitude of the color change is fairly minor compared to that of supernovae, which typically become redder much more rapidly (see Figure 7 of \citet{holoien14b} for a comparison).

Figure~\ref{fig:sed_comp} shows a comparison of the inferred blackbody SEDs for all three objects from epochs shortly after discovery. In all three cases the fit shown used both {\swift} and ground-based data, and for ASASSN-14li and {\name} the data were fit assuming a temperature prior. ASASSN-14ae and {\name} exhibited similar early temperatures of roughly $T\sim20000$~K while ASASSN-14li was hotter, with $T\sim35000$~K. Though {\name} is more luminous and evolved more rapidly than the other two ASAS-SN TDEs, its inferred blackbody temperature is not unusual for an optically-selected TDE.


\begin{figure}
\centering
\includegraphics[width=0.48\textwidth]{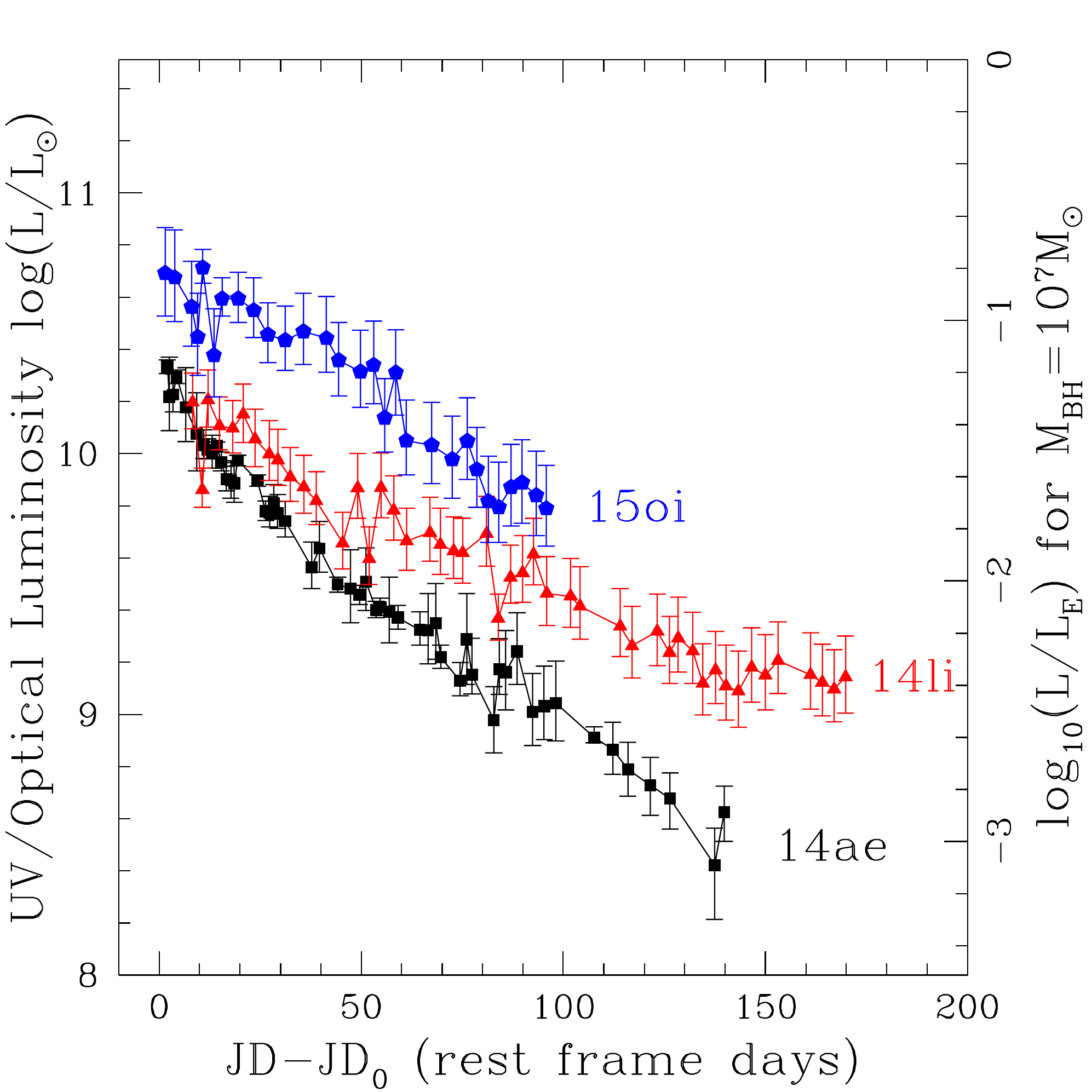}
\caption{Comparison of the evolution of the luminosity inferred from the blackbody SED fits for {\name} (blue pentagons), ASASSN-14ae (black squares), and ASASSN-14li (red triangles). The X-ray emission detected in {\name} and ASASSN-14li is not included in the luminosity measurements. The scale on the right shows the luminosity in units of the Eddington luminosity for a $M_{BH}=10^7$~{\msun} black hole. {\name} is the most luminous of the three TDEs by a clear margin, but its luminosity declines at a rate similar to that of ASASSN-14ae, and the slower-fading ASASSN-14li is of comparable luminosity roughly 70 days after discovery.}
\label{fig:lum_comp}
\end{figure}

\begin{figure}
\centering
\includegraphics[width=0.48\textwidth]{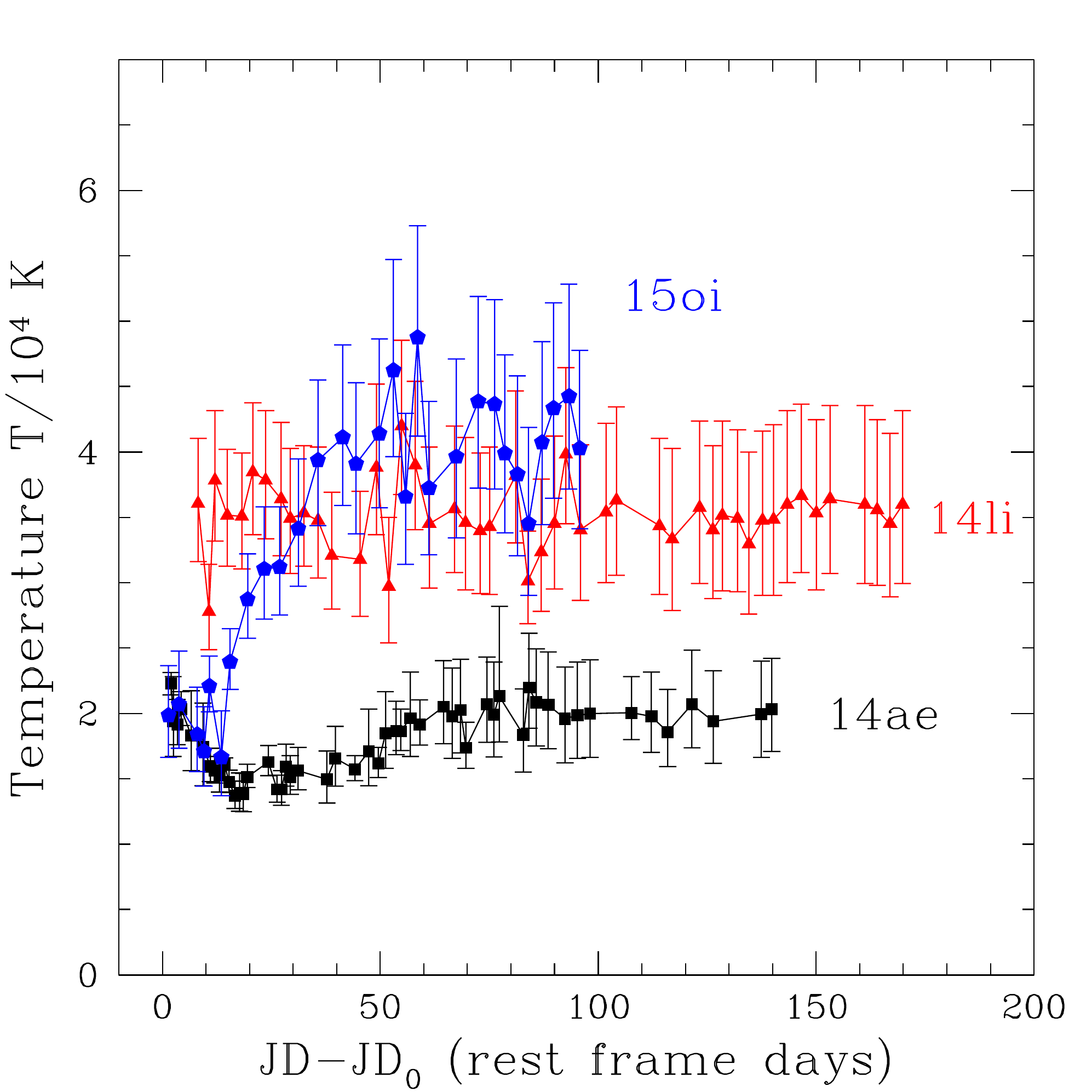}
\caption{Comparison of the blackbody temperature evolution inferred for {\name} (blue pentagons), ASASSN-14ae (black squares), and ASASSN-14li (red triangles). The temperature was fit with a prior for all epochs of {\name} and ASASSN-14li and for those epochs of ASASSN-14ae that did not include {\swift} observations, as described in the text. While ASASSN-14ae and ASASSN-14li exhibit the constant temperature evolution common in TDEs, {\name} shows significant evolution, becoming much hotter than both other ASAS-SN TDEs in the first  $\sim70$ days following discovery, although the maximum temperature is prior-dependent.}
\label{fig:temp_comp}
\end{figure}

\begin{figure}
\centering
\includegraphics[width=0.48\textwidth]{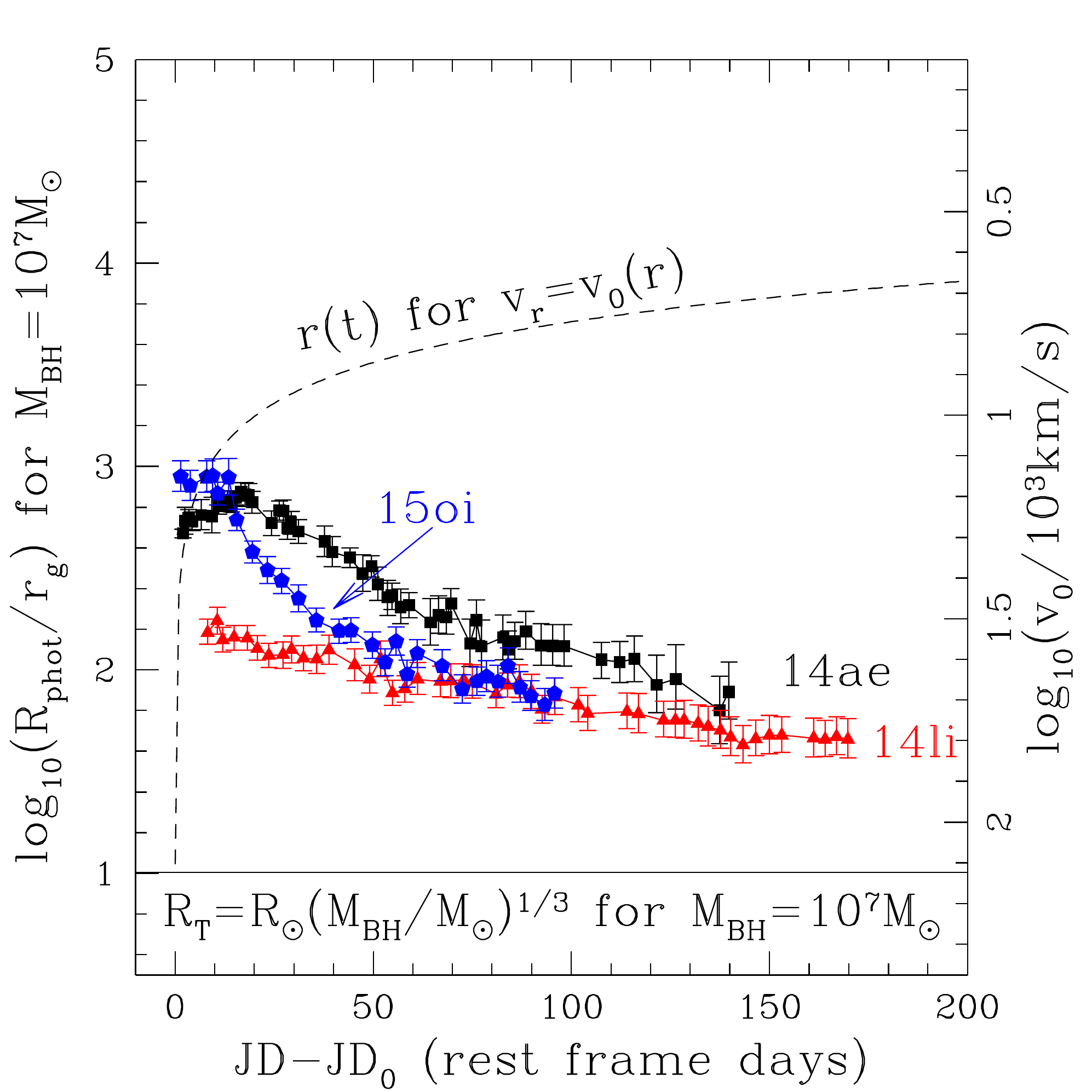}
\caption{Comparison of the evolution of the photospheric radius inferred from the blackbody SED fit for {\name} (blue pentagons), ASASSN-14ae (black squares), and ASASSN-14li (red triangles). The left scale gives the radius in units of the gravitational radius of a $M_{BH}=10^7$~{\msun} black hole, and the right scale converts this into a velocity as $v_0 = c(2 r_g/r)^{1/2}$. For comparison, the horizontal line gives the tidal disruption radius $R_T$ of the Sun, and the dashed line shows the radial evolution $r(t)$ of an orbit expanding at $v_0(r)$. The photospheric radii are roughly bounded by $r(t)$ at peak and then shrink back towards the tidal radius. {\name} shrinks fastest, going from the largest to the most compact in roughly 70 days.}
\label{fig:rad_comp}
\end{figure}

\begin{figure}
\centering
\includegraphics[width=0.48\textwidth]{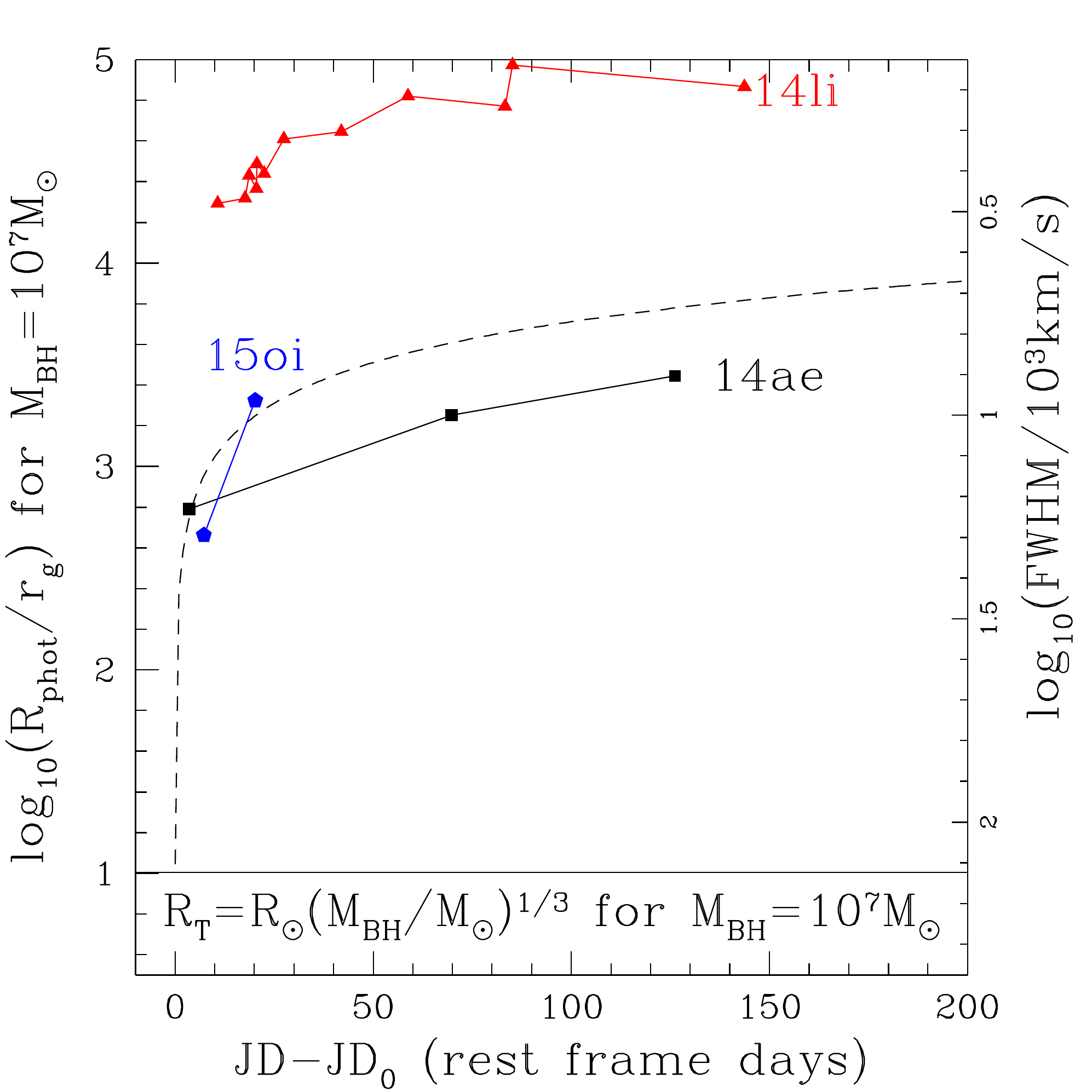}
\caption{Comparison of the evolution of the FWHM (right scale, km~s$^{-1}$).  The left scale shows the radius evolution $r(t)$ assuming a radial velocity of $v_0 = c(2 r_g/r)^{1/2}$.  Also as in Figure~\ref{fig:rad_comp}, the horizontal line gives the tidal disruption radius $R_T$ of the Sun, and the dashed line shows $r(t)$. Results are shown for  the most prominent emission line in each ASAS-SN TDE. H$\alpha$ is shown for ASASSN-14ae (black squares) and ASASSN-14li (red triangles), while \ion{He}{II} 4686\AA~is shown for {\name} (blue pentagons). To the extent that the velocity can be mapped into a radius, the line emission regions lie outside the photospheric radii (Figure~\ref{fig:rad_comp}), close to or beyond $r(t)$, and evolve to larger radii. {\name} exhibits a much more rapid narrowing in its emission lines than ASASSN-14ae or ASASSN-14li.}
\label{fig:vel_comp}
\end{figure}

Finally, we compare the blackbody luminosity, temperature, radius, and emission line evolution inferred for the three ASAS-SN TDEs in Figures~\ref{fig:lum_comp},~\ref{fig:temp_comp}, \ref{fig:rad_comp}, and \ref{fig:vel_comp}, respectively. In all cases the bolometric luminosity was inferred from the optical/UV SED fits, as described in \S\ref{sec:sedanal}, and does not include the X-ray emission detected for {\name} or ASASSN-14li. {\name} is considerably more luminous than the other two TDEs, but its bolometric luminosity falls at a fairly similar rate to that of ASASSN-14ae. ASASSN-14li fades much more slowly than the other two; while it initially has a similar luminosity as ASASSN-14ae, by later epochs it is nearly an order of magnitude more luminous than ASASSN-14ae, and ASASSN-15oi has fallen to a comparable luminosity. This is perhaps correlated with the mass of the black hole involved in the TDE: though the three ASAS-SN TDEs have similar black hole masses given the uncertainties on the mass estimates, ASASSN-14li ostensibly had the least massive black hole, at $M_{BH}\sim10^{6.7}$~{\msun}, while ASASSN-14ae and {\name} had black holes of masses $M_{BH}\sim10^{6.8}$~{\msun} and $M_{BH}\sim10^{7.1}$~{\msun}, respectively.


Including both apparently thermal optical/UV and X-ray emission (where observed), we estimate total radiated energies of $1.7\times10^{50}$~ergs, $7.0\times10^{50}$~ergs, and $6.6\times10^{50}$~ergs for ASASSN-14ae, 14li, and 15oi, respectively.  Converted to an equivalent rest mass, these energies correspond to 0.001, 0.004, and 0.004 {\msun}, compared to a typical TDE stellar mass of $M_* = 0.3 M_{0.3}M_\odot$ \citep{kochanek16}. If the radiated energy is $E = \eta f M_* c^2$ where $\eta \simeq 0.1$ is the radiative efficiency of accretion and $f$ is the fraction of the stellar mass accreted ($f\simeq 0.5$ is bound to the black hole), then we must have $\eta f \simeq$ $0.003 M_{0.3}^{-1}$, $0.013 M_{0.3}^{-1}$ and $0.012 M_{0.3}^{-1}$ for these three events. All three events require either that TDE activity is less radiatively efficient than normal disk accretion ($\eta \ll 0.1$) or that a very small fraction of the bound mass ($f \ll 0.5$) is accreted. Variants of these possibilities appear in recent theoretical models \citep[e.g.,][]{metzger15,piran15,strubbe15,svirski15}.


{\name} stands out very clearly from the other two ASAS-SN TDEs in its temperature evolution. While ASASSN-14ae and ASASSN-14li evolve at nearly constant temperatures, {\name} increases in temperature. The exact increase is poorly constrained because the data do not sample the peak of the SED, but it is consistent with a model where the temperature linearly increases from $T\sim20000$~K to $T\sim40000$~K over roughly 50 days and then remains roughly constant thereafter. {\name} was fit with the temperature prior described in \S\ref{sec:sedanal} above and ASASSN-14li was fit with a prior of $\log T/K=4.55\pm0.05$, as described in \citet{holoien16a}. As the UV data for ASASSN-14ae captured the peak of the transient's SED, its temperature was not fit with a prior for those epochs where {\swift} data were available, and was fit with a prior that tracks the UV data for those epochs where {\swift} data was not available. We emphasize that there are systematic uncertainties in these temperature fits, particularly for {\name} and ASASSN-14li, where the fits were made with very strong priors due to the lack of FUV data, resulting in the lower scatter and larger uncertainties seen in Figure~\ref{fig:temp_comp}. However, these fits are consistent with the data, and the early color evolution of {\name} makes it clear that it has an increasing temperature in early epochs, where the other two ASAS-SN TDEs do not.

Figure~\ref{fig:rad_comp} shows the evolution of the apparent photospheric radii of the ASAS-SN TDEs. All three show the same decreasing trend, but the rate at which the radius falls differs significantly between the three TDEs, with {\name} standing out from the other two events. To put the radius estimates on a more physical scale, Figure~\ref{fig:rad_comp} shows the radii in units of the gravitational radius $r_g=GM_{BH}/c^2 = 1.5 \times 10^{12}$~cm of a $M_{BH} = 10^7 M_\odot$ black hole. The mass estimates of $M_{BH} \sim 10^{6.8}$, $10^{6.7}$ and $10^{7.1}M_\odot$ for ASASSN-14ae, 14li and 15oi are sufficiently close to this mass to use a common scaling given the overall dynamic range.  For comparison, we also show the tidal disruption radius of the Sun, $R_T = R_\odot(M_{BH}/M_\odot)^{1/3} \simeq 1.5 \times 10^{13}~\hbox{cm} \simeq 10 r_g$, and the radius corresponding to expansion at $v_0(r) = c(2 r_g/r)^{1/2}$ and initiated at the origin at $t=0$. To facilitate comparisons to velocities, we also convert the radial scale on the left to the corresponding velocity scale $v_0$ on the right.  


Broadly speaking, the initial photospheric radii are close to the radius that would be achieved by free expansion at $v_0(r)$ from the disruption radius $R_T$ rather than being comparable to $R_T$ or smaller. Given the uncertainties in the black hole masses and the temporal start of the transients, this is mainly a qualitative observation. The apparent photospheres then shrink, but are still well outside $R_T$ at the end of the phase where they can be well-estimated.  That the photospheric radii are all large compared to $R_T$ favors models where the optical/UV emission is reprocessed by debris on large scales rather than direct emission from a disk. \cite{piran15} and \citet{svirski15} propose generating the luminosity by shocks at apocenter, rather than by reprocessing disk emission, but this would imply a much smaller emitting area than $ 4 \pi r^2$ at the apocentric radius.  If the covering fraction of the shock emission region is $f^2$, then the radius has to increase to $r/f$ to produce the observed luminosity and this would initially lie well outside the radius corresponding to an orbit expanding at $v_0(r)$ if $f\sim0.1$, contrary to what is expected for the self-intersection radius of tidal streams associated with a $10^{7}M_\odot$ black hole \citep{dai15}. If the dominant source of line emission was closely associated with the photosphere, we would expect the typical line widths to steadily increase with time. As we noted earlier, this is also the pattern observed in AGN, where line widths increase with decreasing luminosity.

For comparison to the photospheric evolution, Figure~\ref{fig:vel_comp} shows the evolution of the FWHM of the most prominent emission line in each TDE (H$\alpha$ for ASASSN-14ae and ASASSN-14li, and \ion{He}{ii} 4686\AA~for {\name}). The scales and limits of Figure~\ref{fig:vel_comp} are the same as for Figure~\ref{fig:rad_comp} except for relabeling the circular velocity to be the FWHM (i.e. just taking $v_0 = c(2 r_g/r)^{1/2}=\rm{FWHM}$) to allow qualitative comparisons of the radial and velocity scales. For ASASSN-14ae and ASASSN-14li, the initial line formation region is of order the photospheric radius. For ASASSN-14li, the two scales are quite different. The simplest explanation for the mild conflict with expansion at velocity $v_0(r)$ would be to shift the start of the expansion to be 10-20~days earlier. In all three cases, the FWHM diminishes with time, moving outwards in radius if we can interpret the velocities kinematically. At late times, the dominant source of line emission appears to arise from radii far outside the (continuum) photosphere. Though we are only able to obtain two clear measurements of the \ion{He}{ii}~4686\AA~line for {\name}, the data show that the line is narrowing much more rapidly than H$\alpha$ did for the previous two ASAS-SN TDEs.  The primary caveat for the line evolution is that continuum subtraction is challenging for all these events, particularly at late times, and an additional (but sub-dominant) broad emission component might be hard to identify.

The extensive optical and UV data sets obtained for the three ASAS-SN TDEs allows us to perform comparisons between them that are largely impossible with other, fainter TDEs. While the three TDEs show unique characteristics, their colors, temperatures, and luminosity evolution are all broadly similar.


\section{Conclusions}
\label{sec:disc}

{\name} was discovered by ASAS-SN on 2015 August 14 and had a peak absolute $V$-band magnitude of $M_V\sim-20.5$. Follow-up data indicate that it is consistent both photometrically and spectroscopically with previously discovered He-rich, optically-selected TDEs, and is inconsistent with either a supernova or an AGN event. Unlike other optically discovered TDEs, {\name} faded rapidly in the optical and UV, and spectra obtained $\sim3$ months after discovery indicate that the transient features had almost completely disappeared. If we fit the optical and UV emission with a single blackbody model, the best-fit temperature is $T\sim20000$~K, and the temperature becomes hotter in later epochs. The early luminosity declines steadily at a rate best fit by an exponential decay, $L\propto e^{-(t-t_0)/\tau}$, with $\tau\simeq46.5$~days, but is also marginally consistent with a power law decline, $L\propto(t-t_0)^{-5/3}$. Like ASASSN-14li \citep{holoien16a}, {\name} exhibits soft X-ray emission, though this emission is weaker than previous limits and cannot be conclusively tied to the TDE flare. Late-time X-ray observations to see if the X-ray emission starts to fade are needed to confirm the association of the X-ray emission with the transient. {\name} is the third nearby TDE discovered by ASAS-SN, and possibly the second to exhibit both X-ray and optical/UV emission.

Early follow-up spectra show a strong blue continuum and broad \ion{He}{ii} 4686\AA~and \ion{He}{i} 5725\AA~emission features, and while the spectroscopic features evolve rapidly, their evolution is consistent with that of a TDE, and not that of a supernova or quasi-steady state AGN. These spectral features are characteristic of He-rich TDEs like PS1-10jh and PTF09ge \citep{gezari12b,arcavi14}, making {\name} another member of this intriguing class of objects which exhibit strong helium emission lines but no corresponding hydrogen emission. That He-dominated spectra are so common essentially rules out the possibility that they are produced by the disruption of helium stars. While it might be possible to explain a single event such as PS1-10jh with this scenario \citep{gezari12b,strubbe15}, it seems improbable that He stars can represent $\sim1/3$ of the overall optical/UV TDE rate \citep[see][]{kochanek16}. 

In \citet{holoien16a}, we found that a 90\% confidence interval on the implied TDE rate given the ASAS-SN discoveries was $r=(2.2 - 17.0) \times 10^{-5}~{\rm yr}^{-1}$ per galaxy, a range that is consistent with theoretical estimates \citep[e.g.,][]{stone16,kochanek16} while also significantly higher but consistent with the rates inferred from other optical surveys \citep[e.g.,][]{velzen14}, given the uncertainties. The discovery of a third ASAS-SN TDE less than a year after the previous one implies that the relatively high (compared to previous observational studies) rate from \citet{holoien16a} is correct, which helps to alleviate some of the tension between observed and theoretical rates. Including this third TDE, ASAS-SN is finding roughly 1 TDE for every 70 type Ia supernovae (3 and 211, respectively, at the time of writing), a rate that is significantly higher than that of other surveys (see \citet{holoien16a} for a detailed comparison). This suggests that the ASAS-SN TDE sample is significantly more complete than previous surveys, and consequently that our rate estimate may be more accurate.

ASAS-SN has now discovered the three nearest optically discovered TDEs, each of which has been extensively observed over long periods of time in the optical, UV, and X-rays. Due to the nearby and bright nature of objects discovered by ASAS-SN, future TDEs will be similarly easy to observe with a wide variety of instruments, which will allow the creation of a catalog of well-studied TDEs that can be used to study the early and late-time behaviors of these transients, which cannot be done with higher-redshift objects. Moreover, due to the rapid evolution of the black hole mass function, the TDE rate should decline rapidly with increasing redshift \citep{kochanek16}, making TDEs creatures of the local universe probed by ASAS-SN. Given the success of the ASAS-SN TDE search to-date, we expect it will remain a powerful tool for finding and studying TDEs and other bright transients in the future.

\section*{Acknowledgments}

The authors thank PI Neil Gehrels and the {\swift} ToO team for promptly approving and executing our observations. We thank LCOGT and its staff for their continued support of ASAS-SN.

ASAS-SN is supported by NSF grant AST-1515927. Development of ASAS-SN has been supported by NSF grant AST-0908816, the Center for Cosmology and AstroParticle Physics at the Ohio State University, the Mt. Cuba Astronomical Foundation, and by George Skestos.

TW-SH is supported by the DOE Computational Science Graduate Fellowship, grant number DE-FG02-97ER25308. Support for JLP is in part provided by FONDECYT through the grant 1151445 and by the Ministry of Economy, Development, and Tourism's Millennium Science Initiative through grant IC120009, awarded to The Millennium Institute of Astrophysics, MAS. PC and SD are supported by ``the Strategic Priority Research Program-The Emergence of Cosmological Structures'' of the Chinese Academy of Sciences (Grant No. XDB09000000) and Project 11573003 supported by NSFC. BJS is supported by NASA through Hubble Fellowship grant HST-HF-51348.001 awarded by the Space Telescope Science Institute, which is operated by the Association of Universities for Research in Astronomy, Inc., for NASA, under contract NAS 5-26555. JFB is supported by NSF grant PHY-1404311. EJ acknowledges support from the Marie Curie Actions of the European Commission (FP7-COFUND).

This research has made use of the XRT Data Analysis Software (XRTDAS) developed under the responsibility of the ASI Science Data Center (ASDC), Italy. At Penn State the NASA {\swift} program is support through contract NAS5-00136.

This research uses data obtained through the Telescope Access Program (TAP), which has been funded by the ``Strategic Priority Research Program--The Emergence of Cosmological Structures'' of the Chinese Academy of Sciences (Grant No. XDB09000000) and the Special Fund for Astronomy from the Ministry of Finance.

This research was made possible through the use of the AAVSO Photometric All-Sky Survey (APASS), funded by the Robert Martin Ayers Sciences Fund.

This research has made use of data provided by Astrometry.net \citep{barron08}.

This paper uses data products produced by the OIR Telescope Data Center, supported by the Smithsonian Astrophysical Observatory.

This publication makes use of data products from the Two Micron All Sky Survey, which is a joint project of the University of Massachusetts and the Infrared Processing and Analysis Center/California Institute of Technology, funded by NASA and the NSF.

This publication makes use of data products from the Wide-field Infrared Survey Explorer, a joint project of the University of California, Los Angeles, and the Jet Propulsion Laboratory/California Institute of Technology, funded by the National Aeronautics and Space Administration.

This research has made use of the NASA/IPAC Extragalactic Database (NED), which is operated by the Jet Propulsion Laboratory, California Institute of Technology, under contract with the National Aeronautics and Space Administration.

This work is based in part on observations collected at the European Organisation for Astronomical Research in the Southern Hemisphere, Chile as part of PESSTO, (the Public ESO Spectroscopic Survey for Transient Objects Survey) ESO program 188.D-3003, 191.D-0935.

This publication makes use of data obtained from the Weizmann interactive supernova data repository \citep[WISEREP;][]{yaron12}.

\bibliographystyle{mn2e}
\bibliography{bibliography}


\appendix
\section{Follow-up Photometry}
All follow-up photometry are presented in Table~\ref{tab:phot} and Table~\ref{tab:xray}, while information about follow-up spectroscopic observations is given in Table~\ref{tab:spectra}. All optical and UV photometry is presented in the Vega system.


\begin{table*}
\begin{minipage}{\textwidth}
\centering
\caption{Photometric data of {\name}.\hfill}
\renewcommand{\arraystretch}{1.2}
\begin{tabular}{cccc|cccc}
\hline
MJD & Magnitude &  Filter & Telescope & MJD & Magnitude &  Filter & Telescope\\
\hline
57249.532 & 15.45 0.129 & $I$ & LCOGT & 57327.129 & 17.26 0.045 & $V$ & LCOGT\\ 
57252.002 & 15.54 0.158 & $I$ & LCOGT & 57329.844 & 17.25 0.055 & $V$ & LCOGT\\ 
57256.428 & 15.46 0.133 & $I$ & LCOGT & 57332.791 & 17.28 0.051 & $V$ & LCOGT\\ 
57258.008 & 15.43 0.143 & $I$ & LCOGT & 57335.816 & 17.29 0.065 & $V$ & LCOGT\\ 
57262.251 & 15.34 0.193 & $I$ & LCOGT & 57344.415 & 17.22 0.060 & $V$ & LCOGT\\ 
57264.083 & 15.83 0.150 & $I$ & LCOGT & 57346.420 & 17.17 0.066 & $V$ & LCOGT\\ 
57268.052 & 15.72 0.129 & $I$ & LCOGT & 57349.423 & 17.32 0.056 & $V$ & LCOGT\\ 
57269.908 & 15.71 0.133 & $I$ & LCOGT & 57355.035 & 17.25 0.052 & $V$ & LCOGT\\ 
57271.882 & 15.70 0.143 & $I$ & LCOGT & 57259.334 & 16.12 0.071 & $V$ & {\swift}\\ 
57273.924 & 15.80 0.148 & $I$ & LCOGT & 57264.392 & 16.42 0.071 & $V$ & {\swift}\\ 
57276.874 & 15.80 0.142 & $I$ & LCOGT & 57268.788 & 16.88 0.1 & $V$ & {\swift}\\ 
57279.895 & 15.81 0.140 & $I$ & LCOGT & 57272.514 & 16.86 0.091 & $V$ & {\swift}\\ 
57282.476 & 15.85 0.133 & $I$ & LCOGT & 57276.096 & 16.98 0.1 & $V$ & {\swift}\\ 
57291.383 & 15.98 0.154 & $I$ & LCOGT & 57280.688 & 17.03 0.091 & $V$ & {\swift}\\ 
57294.522 & 15.99 0.141 & $I$ & LCOGT & 57285.415 & 16.99 0.17 & $V$ & {\swift}\\ 
57299.495 & 15.80 0.139 & $I$ & LCOGT & 57291.590 & 17.01 0.13 & $V$ & {\swift}\\ 
57302.481 & 15.95 0.200 & $I$ & LCOGT & 57294.588 & 17.06 0.13 & $V$ & {\swift}\\ 
57304.921 & 15.86 0.262 & $I$ & LCOGT & 57300.236 & 17.47 0.16 & $V$ & {\swift}\\ 
57307.106 & 15.93 0.126 & $I$ & LCOGT & 57303.628 & 17.29 0.15 & $V$ & {\swift}\\ 
57307.140 & 15.96 0.125 & $I$ & LCOGT & 57306.494 & 17.08 0.12 & $V$ & {\swift}\\ 
57316.018 & 15.98 0.134 & $I$ & LCOGT & 57309.413 & 17.54 0.22 & $V$ & {\swift}\\ 
57319.059 & 16.05 0.152 & $I$ & LCOGT & 57312.137 & 17.09 0.13 & $V$ & {\swift}\\ 
57321.796 & 16.04 0.116 & $I$ & LCOGT & 57318.671 & 17.28 0.25 & $V$ & {\swift}\\ 
57327.133 & 16.03 0.126 & $I$ & LCOGT & 57324.051 & 17.82 0.36 & $V$ & {\swift}\\ 
57329.849 & 15.88 0.128 & $I$ & LCOGT & 57327.961 & 17.16 0.15 & $V$ & {\swift}\\ 
57332.796 & 16.02 0.159 & $I$ & LCOGT & 57330.420 & 17.26 0.25 & $V$ & {\swift}\\ 
57335.821 & 16.06 0.157 & $I$ & LCOGT & 57333.480 & 17.30 0.2 & $V$ & {\swift}\\ 
57346.424 & 15.83 0.140 & $I$ & LCOGT & 57336.141 & 17.61 0.22 & $V$ & {\swift}\\ 
57349.419 & 15.87 0.133 & $I$ & LCOGT & 57339.334 & 16.84 0.21 & $V$ & {\swift}\\ 
57355.039 & 15.92 0.121 & $I$ & LCOGT & 57342.069 & 16.88 0.18 & $V$ & {\swift}\\ 
57249.533 & 16.01 0.053 & $V$ & LCOGT & 57345.793 & 17.12 0.16 & $V$ & {\swift}\\ 
57252.004 & 16.06 0.055 & $V$ & LCOGT & 57348.384 & 17.23 0.16 & $V$ & {\swift}\\ 
57256.429 & 16.1 0.053 & $V$ & LCOGT & 57249.531 & 16.09 0.055 & $B$ & LCOGT\\ 
57258.009 & 16.3 0.225 & $V$ & LCOGT & 57252.001 & 16.2 0.075 & $B$ & LCOGT\\ 
57262.252 & 16.52 0.144 & $V$ & LCOGT & 57256.426 & 16.34 0.042 & $B$ & LCOGT\\ 
57266.065 & 16.55 0.112 & $V$ & LCOGT & 57258.007 & 16.44 0.065 & $B$ & LCOGT\\ 
57268.053 & 16.63 0.047 & $V$ & LCOGT & 57262.250 & 16.52 0.082 & $B$ & LCOGT\\ 
57269.909 & 16.76 0.050 & $V$ & LCOGT & 57264.081 & 16.58 0.062 & $B$ & LCOGT\\ 
57271.883 & 16.79 0.062 & $V$ & LCOGT & 57266.062 & 16.87 0.063 & $B$ & LCOGT\\ 
57273.926 & 16.83 0.058 & $V$ & LCOGT & 57268.051 & 16.99 0.068 & $B$ & LCOGT\\ 
57276.875 & 16.93 0.057 & $V$ & LCOGT & 57269.906 & 17.05 0.060 & $B$ & LCOGT\\ 
57278.172 & 16.94 0.062 & $V$ & LCOGT & 57271.880 & 17.19 0.074 & $B$ & LCOGT\\ 
57282.475 & 16.98 0.050 & $V$ & LCOGT & 57273.923 & 17.38 0.067 & $B$ & LCOGT\\ 
57285.424 & 16.9 0.130 & $V$ & LCOGT & 57276.872 & 17.26 0.069 & $B$ & LCOGT\\ 
57291.382 & 17.03 0.081 & $V$ & LCOGT & 57278.169 & 17.42 0.076 & $B$ & LCOGT\\ 
57294.521 & 17.13 0.074 & $V$ & LCOGT & 57282.474 & 17.47 0.065 & $B$ & LCOGT\\ 
57304.919 & 17.22 0.074 & $V$ & LCOGT & 57285.423 & 17.62 0.083 & $B$ & LCOGT\\ 
57307.104 & 17.18 0.047 & $V$ & LCOGT & 57291.381 & 17.7 0.115 & $B$ & LCOGT\\ 
57307.138 & 17.18 0.056 & $V$ & LCOGT & 57294.520 & 17.58 0.242 & $B$ & LCOGT\\ 
57312.170 & 17.2 0.068 & $V$ & LCOGT & 57295.130 & 17.67 0.085 & $B$ & LCOGT\\ 
57315.877 & 17.28 0.082 & $V$ & LCOGT & 57299.493 & 17.95 0.098 & $B$ & LCOGT\\ 
57315.894 & 17.14 0.073 & $V$ & LCOGT & 57304.918 & 17.59 0.086 & $B$ & LCOGT\\ 
57316.016 & 17.24 0.075 & $V$ & LCOGT & 57307.137 & 17.93 0.067 & $B$ & LCOGT\\ 
57319.057 & 17.23 0.059 & $V$ & LCOGT & 57312.168 & 17.98 0.077 & $B$ & LCOGT\\ 
57321.794 & 17.14 0.077 & $V$ & LCOGT & 57315.873 & 18.27 0.130 & $B$ & LCOGT\\
\hline
\end{tabular}
\label{tab:phot}
\end{minipage}
\end{table*}

\begin{table*}
\begin{minipage}{\textwidth}
\centering
\renewcommand{\arraystretch}{1.2}
\begin{tabular}{cccc|cccc}
\hline
MJD & Magnitude &  Filter & Telescope & MJD & Magnitude &  Filter & Telescope\\
\hline
57316.015 & 18.09 0.086 & $B$ & LCOGT & 57339.331 & 17.81 0.271 & $U$ & {\swift}\\ 
57319.055 & 18.21 0.108 & $B$ & LCOGT & 57342.066 & 17.84 0.231 & $U$ & {\swift}\\ 
57321.792 & 18.45 0.179 & $B$ & LCOGT & 57345.787 & 18.08 0.201 & $U$ & {\swift}\\ 
57327.124 & 18.12 0.058 & $B$ & LCOGT & 57348.377 & 17.82 0.161 & $U$ & {\swift}\\ 
57329.839 & 18.14 0.049 & $B$ & LCOGT & 57259.325 & 14.60 0.042 & $W1$ & {\swift}\\ 
57332.786 & 18.31 0.099 & $B$ & LCOGT & 57264.379 & 15.02 0.042 & $W1$ & {\swift}\\ 
57335.812 & 18.18 0.102 & $B$ & LCOGT & 57268.719 & 15.30 0.05 & $W1$ & {\swift}\\ 
57344.411 & 18.25 0.077 & $B$ & LCOGT & 57272.503 & 15.55 0.05 & $W1$ & {\swift}\\ 
57346.415 & 18.11 0.131 & $B$ & LCOGT & 57276.085 & 15.89 0.058 & $W1$ & {\swift}\\ 
57349.414 & 18.27 0.076 & $B$ & LCOGT & 57280.675 & 16.07 0.058 & $W1$ & {\swift}\\ 
57353.420 & 18.24 0.064 & $B$ & LCOGT & 57285.411 & 16.24 0.076 & $W1$ & {\swift}\\ 
57355.030 & 18.35 0.071 & $B$ & LCOGT & 57291.384 & 16.52 0.067 & $W1$ & {\swift}\\ 
57259.329 & 16.33 0.045 & $B$ & {\swift} & 57294.579 & 16.66 0.067 & $W1$ & {\swift}\\ 
57264.384 & 16.73 0.045 & $B$ & {\swift} & 57297.048 & 16.91 0.076 & $W1$ & {\swift}\\ 
57268.724 & 16.95 0.054 & $B$ & {\swift} & 57300.226 & 16.91 0.076 & $W1$ & {\swift}\\ 
57272.507 & 17.26 0.063 & $B$ & {\swift} & 57303.618 & 17.07 0.076 & $W1$ & {\swift}\\ 
57276.089 & 17.33 0.063 & $B$ & {\swift} & 57306.483 & 17.14 0.076 & $W1$ & {\swift}\\ 
57280.679 & 17.46 0.063 & $B$ & {\swift} & 57309.407 & 17.16 0.104 & $W1$ & {\swift}\\ 
57285.413 & 17.57 0.122 & $B$ & {\swift} & 57312.128 & 17.41 0.095 & $W1$ & {\swift}\\ 
57291.388 & 17.67 0.092 & $B$ & {\swift} & 57318.666 & 17.55 0.133 & $W1$ & {\swift}\\ 
57294.582 & 17.68 0.092 & $B$ & {\swift} & 57324.047 & 17.98 0.192 & $W1$ & {\swift}\\ 
57300.230 & 17.71 0.092 & $B$ & {\swift} & 57327.909 & 17.68 0.143 & $W1$ & {\swift}\\ 
57303.622 & 17.80 0.102 & $B$ & {\swift} & 57330.416 & 17.63 0.163 & $W1$ & {\swift}\\ 
57306.487 & 17.68 0.092 & $B$ & {\swift} & 57333.412 & 17.60 0.153 & $W1$ & {\swift}\\ 
57309.409 & 17.80 0.132 & $B$ & {\swift} & 57336.132 & 17.89 0.133 & $W1$ & {\swift}\\ 
57312.132 & 17.68 0.102 & $B$ & {\swift} & 57339.330 & 17.72 0.192 & $W1$ & {\swift}\\ 
57318.668 & 18.13 0.181 & $B$ & {\swift} & 57342.064 & 18.06 0.202 & $W1$ & {\swift}\\ 
57324.049 & 17.91 0.171 & $B$ & {\swift} & 57345.784 & 18.26 0.173 & $W1$ & {\swift}\\ 
57327.911 & 18.44 0.231 & $B$ & {\swift} & 57348.374 & 18.11 0.143 & $W1$ & {\swift}\\ 
57330.418 & 18.08 0.221 & $B$ & {\swift} & 57259.335 & 14.44 0.042 & $M2$ & {\swift}\\ 
57333.414 & 18.08 0.211 & $B$ & {\swift} & 57264.394 & 14.87 0.042 & $M2$ & {\swift}\\ 
57336.135 & 18.17 0.181 & $B$ & {\swift} & 57268.789 & 15.21 0.05 & $M2$ & {\swift}\\ 
57342.066 & 17.84 0.171 & $B$ & {\swift} & 57272.515 & 15.36 0.05 & $M2$ & {\swift}\\ 
57345.788 & 18.26 0.171 & $B$ & {\swift} & 57276.098 & 15.63 0.05 & $M2$ & {\swift}\\ 
57348.378 & 18.27 0.171 & $B$ & {\swift} & 57280.689 & 15.79 0.05 & $M2$ & {\swift}\\ 
57259.328 & 15.25 0.045 & $U$ & {\swift} & 57285.416 & 15.97 0.067 & $M2$ & {\swift}\\ 
57264.383 & 15.69 0.045 & $U$ & {\swift} & 57291.591 & 16.18 0.05 & $M2$ & {\swift}\\ 
57268.722 & 15.99 0.045 & $U$ & {\swift} & 57294.589 & 16.38 0.058 & $M2$ & {\swift}\\ 
57272.506 & 16.19 0.054 & $U$ & {\swift} & 57300.237 & 16.67 0.058 & $M2$ & {\swift}\\ 
57276.088 & 16.51 0.063 & $U$ & {\swift} & 57303.629 & 16.81 0.058 & $M2$ & {\swift}\\ 
57280.678 & 16.73 0.063 & $U$ & {\swift} & 57306.495 & 16.93 0.058 & $M2$ & {\swift}\\ 
57285.412 & 16.99 0.122 & $U$ & {\swift} & 57309.413 & 17.00 0.076 & $M2$ & {\swift}\\ 
57291.387 & 16.96 0.082 & $U$ & {\swift} & 57312.138 & 17.07 0.067 & $M2$ & {\swift}\\ 
57294.581 & 17.26 0.102 & $U$ & {\swift} & 57318.869 & 17.40 0.182 & $M2$ & {\swift}\\ 
57300.229 & 17.36 0.102 & $U$ & {\swift} & 57324.052 & 17.59 0.114 & $M2$ & {\swift}\\ 
57303.621 & 17.47 0.112 & $U$ & {\swift} & 57327.962 & 17.56 0.076 & $M2$ & {\swift}\\ 
57306.486 & 17.45 0.102 & $U$ & {\swift} & 57330.421 & 17.41 0.114 & $M2$ & {\swift}\\ 
57309.408 & 17.67 0.161 & $U$ & {\swift} & 57333.481 & 17.86 0.104 & $M2$ & {\swift}\\ 
57312.131 & 17.79 0.151 & $U$ & {\swift} & 57336.142 & 17.82 0.085 & $M2$ & {\swift}\\ 
57318.667 & 17.46 0.161 & $U$ & {\swift} & 57339.334 & 17.84 0.143 & $M2$ & {\swift}\\ 
57324.048 & 17.75 0.211 & $U$ & {\swift} & 57342.069 & 17.84 0.124 & $M2$ & {\swift}\\ 
57327.910 & 17.68 0.181 & $U$ & {\swift} & 57345.794 & 17.94 0.242 & $M2$ & {\swift}\\ 
57330.418 & 17.68 0.231 & $U$ & {\swift} & 57348.385 & 17.92 0.095 & $M2$ & {\swift}\\ 
57333.413 & 17.79 0.221 & $U$ & {\swift} & 57257.794 & 14.77 0.042 & $W2$ & {\swift}\\ 
57336.134 & 17.47 0.141 & $U$ & {\swift} & 57259.33 & 14.83 0.042 & $W2$ & {\swift}\\ 
\hline
\end{tabular}
\end{minipage}
\end{table*}

\begin{table*}
\begin{minipage}{\textwidth}
\centering
\renewcommand{\arraystretch}{1.2}
\begin{tabular}{cccc|cccc}
\hline
MJD & Magnitude &  Filter & Telescope & MJD & Magnitude &  Filter & Telescope\\
\hline
57268.725 & 15.21 0.076 & $W2$ & {\swift} & 57312.133 & 16.79 0.058 & $W2$ & {\swift}\\ 
57272.508 & 15.48 0.042 & $W2$ & {\swift} & 57318.668 & 16.94 0.085 & $W2$ & {\swift}\\ 
57276.091 & 15.68 0.042 & $W2$ & {\swift} & 57324.049 & 17.16 0.095 & $W2$ & {\swift}\\ 
57280.681 & 15.91 0.042 & $W2$ & {\swift} & 57327.911 & 16.94 0.143 & $W2$ & {\swift}\\ 
57285.413 & 16.01 0.058 & $W2$ & {\swift} & 57330.418 & 17.42 0.114 & $W2$ & {\swift}\\ 
57291.389 & 16.12 0.05 & $W2$ & {\swift} & 57336.136 & 17.29 0.076 & $W2$ & {\swift}\\ 
57294.583 & 16.17 0.05 & $W2$ & {\swift} & 57339.332 & 17.47 0.124 & $W2$ & {\swift}\\ 
57300.231 & 16.35 0.05 & $W2$ & {\swift} & 57342.067 & 17.39 0.104 & $W2$ & {\swift}\\ 
57303.623 & 16.42 0.058 & $W2$ & {\swift} & 57345.789 & 17.51 0.085 & $W2$ & {\swift}\\ 
57306.488 & 16.57 0.058 & $W2$ & {\swift} & 57348.379 & 17.60 0.085 & $W2$ & {\swift}\\ 
57309.41 & 16.56 0.067 & $W2$ & {\swift} &  &   &  & \\ 
\hline
\end{tabular}

\medskip
\raggedright
\noindent All magnitudes and uncertainties are presented in the Vega system. Uncertainties are given next to the magnitude measurements. Data are not corrected for Galactic extinction.
\end{minipage}
\end{table*}


\begin{table*}
\begin{minipage}{\textwidth}
\centering
\caption{Swift XRT photometry of {\name}.}
\renewcommand{\arraystretch}{1.2}
\begin{tabular}{ccccc}
\hline
MJD & Flux  & Lower Uncertainty & Upper Uncertainty & Epochs Combined \\
\hline
57263.0 & 4.18 & 2.04 & 1.73 & $01-04$ \\  
57279.0 & 6.80 & 2.04 & 1.73 & $05-09$ \\
57296.0 & 6.77 & 2.31 & 1.97 & $10-13$ \\
57306.0 & 8.32 & 2.98 & 2.35 & $14-16$ \\
57319.0 & 6.80 & 2.25 & 1.88 & $17-20$ \\
57335.0 & 7.79 & 2.20 & 1.88 & $21-24$ \\
57345.0 & 6.17 & 2.98 & 2.35 & $25-27$ \\
\hline
\end{tabular}

\medskip
\raggedright
\noindent All X-ray fluxes and uncertainties are given in units of $10^{-14}$ ergs~s$^{-1}$~cm$^{-2}$. Data are not corrected for Galactic extinction. The ``Epochs Combined'' column indicates which {\swift} epochs were combined to obtain the listed measurement.
\label{tab:xray}
\end{minipage}
\end{table*}


\begin{table*}
\begin{minipage}{\textwidth}
\centering
\caption{Spectroscopic Observations of {\name}.}
\renewcommand{\arraystretch}{1.2}
\begin{tabular}{lclr}
\hline
\multicolumn{1}{c}{UT Date} & MJD & \multicolumn{1}{c}{Telescope/Instrument} & \multicolumn{1}{c}{Exposure (s)} \\
\hline
2015 September 04.22 & 57269.22 & du Pont-2.5m/WFCCD & $1\times900$ \\
2015 September 12.20 & 57277.20 & MDM-2.4m/OSMOS & $1\times600$ \\
2015 October 11.82 & 57306.82 & Tillinghast-1.5m/FAST & $1\times2400$ \\
2015 October 12.82 & 57307.82 & Tillinghast-1.5m/FAST & $1\times2100$ \\
2015 October 15.75 & 57310.75 & Tillinghast-1.5m/FAST & $1\times1320$ \\
2015 October 17.12 & 57312.12 & du Pont-2.5m/WFCCD & $2\times900$ \\
2015 November 07.06 & 57333.06 & du Pont-2.5m/WFCCD & $2\times1200$ \\
\hline
\end{tabular}
\label{tab:spectra}
\end{minipage}
\end{table*}

\end{document}